\documentclass[twoside,twocolumn,aps,prb,superscriptaddress,longbibliography]{revtex4-2}
\usepackage[latin9]{inputenc}
\usepackage{array}
\usepackage{mathrsfs}
\usepackage{multirow}
\usepackage{amsmath}
\usepackage{amssymb}
\usepackage{graphicx}
\PassOptionsToPackage{version=3}{mhchem}
\usepackage{mhchem}

\makeatletter

\providecommand{\tabularnewline}{\\}

\usepackage{float}
\usepackage{soul}
\usepackage{changes}
\usepackage{mathrsfs}
\usepackage{comment}
\usepackage{multirow}
\usepackage{xcolor}

\makeatother

\begin{document}
\title{\textit{Ab initio} studies of influence of periodic-direction electric fields on spin lifetime and spin diffusion length and the validation of an \textit{ab initio} matrix-drift-diffusion model}
\setcounter{page}{1}
\date{\today}
\author{Junqing Xu}
\email{jqxu@hfut.edu.cn}
\affiliation{Department of Physics, Hefei University of Technology, Hefei, Anhui 230601, China}
\author{Can Liu}
\affiliation{Department of Physics, Hefei University of Technology, Hefei, Anhui 230601, China}
\author{Weiwei Chen}
\email{chenweiwei@hfut.edu.cn}
\affiliation{Department of Physics, Hefei University of Technology, Hefei, Anhui 230601, China}
\begin{abstract}
Recently, we developed an \textit{ab initio} approach of spin lifetime
($\tau_{s}$) and spin diffusion length ($l_{s}$) in solids {[}Phys.
Rev. Lett. 135, 046705 (2025){]}, based on a density-matrix master
equation with quantum treatment of electron scattering processes.
In this work, we extend the method to include the drift term due to
an electric field along a periodic direction, implemented using a
Wannier-representation-based covariant derivative. We employ this
approach to investigate the electric-field effect on $\tau_{s}$ and
$l_{s}$ of monolayer WSe$_{2}$, bulk GaAs, bulk GaN, and graphene-$h$-BN
heterostructure. We find that an electric field reduces $\tau_{s}$
of GaAs, due to the induced D\textquoteright yakonov-Perel\textquoteright -type
spin relaxation. In GaN and graphene-$h$-BN, $\tau_{s}$ is significantly
affected, partly because the electric field generates an effective
magnetic field corresponding to the $k$-derivative of Rashba spin-orbit
(magnetic) field. Our results show that $l_{s}$ can be significantly
enhanced or suppressed by a moderate downstream or upstream field
respectively. While the standard drift-diffusion model performs well
for WSe$_{2}$, it can introduce large errors of the electric-field-induced
changes of $l_{s}$ in GaAs, GaN and graphene-$h$-BN. Our proposed
\textit{ab initio} matrix-drift-diffusion model improves results
for GaAs and GaN, but still fails for graphene-$h$-BN. Thus, to accurately
capture the influence of electric fields on $l_{s}$ in realistic
materials, it is necessary to go beyond the drift-diffusion model
and adopt a microscopic \textit{ab initio} methodology. Moreover,
in graphene-$h$-BN, we find that the field-induced changes of $\tau_{s}$
and $l_{s}$ are not only governed by the drift term in the master
equation, but are also significantly affected by the electric-field
modification of the equilibrium density matrix away from Fermi-Dirac
distribution function.
\end{abstract}
\maketitle

\section{Introduction}

Spin lifetime ($\tau_{s}$) and spin diffusion length ($l_{s}$) are
key parameters for spintronic device applications, which aim to achieve
the next generation of low-power electronics by making use of the
spin degree of freedom\citep{vzutic2004spintronics,wu2010spin,avsar2020colloquium}.
In spintronic devices, sufficiently long $\tau_{s}$ and $l_{s}$
are typically required within the spin transport channel to ensure
stable spin detection and manipulation. Accurate simulations of $\tau_{s}$
and $l_{s}$ are invaluable to predict promising spintronic materials
with long enough $\tau_{s}$ and $l_{s}$ and optimizing device operating
conditions. Recently, we developed an \textit{ab initio} approach
of both $\tau_{s}$ and $l_{s}$ in solids, based on a density-matrix
master equation with quantum treatment of electron scattering processes\citep{xu2025predicting}.
It has been successfully applied to disparate materials--including
2D and 3D semiconductors, graphene-$h$-BN heterostructure, the heavy
metal Pt and the antiferromagnet RuO$_{2}$--yielding theoretical
results in good agreement with experimental measurements\citep{li2021valley,dzhioev2004suppression,avsar2020colloquium,li2025spintronics,wesselink2019calculating,bose2022tilted}.

While our approach can readily incorporate an electric field (${\bf E}$)
along a non-periodic direction at the density-functional-theory (DFT)
level\citep{xu2021giant,xu2023substrate}, the treatment of a field
along a periodic direction remained challenging. This capability is
crucial for applications, as a longitudinal ${\bf E}$--essential
for driving charge carriers and their associated spins in devices--can
significantly modulate $l_{s}$ via spin-drift effect\citep{wu2010spin,yu2002electric,yu2002spin,hruvska2006effects,jozsa2008electronic,zucchetti2022electric,miah2008drift}.
Consequently, accurate simulation of this effect is critical for the
predictive modeling of spintronic devices.

Spin-drift effect has been investigated both theoretically and experimentally
by several groups. In semiconductors, its influence on $l_{s}$ is
often estimated using the following drift-diffusion model: $D\nabla^{2}S\left({\bf r}\right)$$\pm$$\mu{\bf E}\cdot\nabla S\left({\bf r}\right)$$=$$S\left({\bf r}\right)/\tau_{s}$,
with $\pm$ corresponding to electrons or holes. $\mu$ is carrier
mobility. $D$ is diffusion coefficient. By solving this equation
and defining $\eta$=$\mu/D$, the following model formula of $l_{s}\left(E\right)$
is obtained\citep{miah2008drift,wu2010spin,yu2002electric}
\begin{align}
\frac{1}{l_{s}\left(E\right)}= & \sqrt{\frac{1}{l_{s0}^{2}}+\frac{E^{2}}{4}\frac{\mu^{2}}{D^{2}}}-\theta\frac{\left|E\right|}{2}\frac{\mu}{D},\label{eq:lsE_from_standard_DD}
\end{align}
where $l_{s0}$ is the zero-electric-field spin diffusion length.
$\theta$ is $\pm$ for downstream or upstream fields respectively.
However, as pointed out in our recent \textit{ab initio} work, the
drift-diffusion model may cause significant errors of $l_{s}$ in
specific systems such as graphene-$h$-BN and wurtzite GaN, where
spatial spin precession substantially influences spin diffusion. It
is therefore essential to assess the validity of this model in such
systems through direct comparison with \textit{ab initio} results.
Other theoretical studies have employed more sophisticated transport
equations\citep{wu2010spin} to simulate spin-drift effect on $l_{s}$.
Nonetheless, these works typically rely on model Hamiltonians and
simplified treatments of scattering processes, and thus are lack of
predictive accuracy and broad applicability.

The effect of a periodic-direction electric field on $\tau_{s}$ was
also studied experimentally and theoretically\citep{wu2010spin,jiang2009electron},
and was found important to materials, in which spin relaxation is
dominated by D\textquoteright yakonov-Perel\textquoteright{} mechanism.
Previous theoretical studies have relied on model Hamiltonians, so
that fully \textit{ab initio} simulations are valuable to improve
the understandings of the electric-field effects.

Here we extend our recently developed \textit{ab initio} approach
of $\tau_{s}$ and $l_{s}$ to include the drift term due to a uniform
dc electric field along a periodic direction, using a covariant derivative
formalism. The extension enables accurate simulation of the electric-field
effect on $\tau_{s}$ and $l_{s}$ across a wide range of materials,
thereby fully addressing the above issues.

The paper is organized as follows. In Sec. II A, we derive the linearized
quantum master equation for electron dynamics and transport under
finite ${\bf E}$. We then describe how to simulate $\tau_{s}\left(E\right)$
and $l_{s}\left(E\right)$ via solving full eigenvalue problems (EVPs).
Section II B introduces several approximate methods: first, calculations
based on Rayleigh-Ritz (RR) method; second, an \textit{ab initio}
matrix-drift-diffusion (ab-mDD) model; and third, three model formulas
of $l_{s}\left(E\right)$ from the standard drift-diffusion model.
In Sec. III, we present theoretical results of $\tau_{s}\left(E\right)$
and $l_{s}\left(E\right)$ by our \textit{ab initio} approach, compared
with estimates from approximate methods, for monolayer WSe$_{2}$,
bulk GaAs, bulk GaN and graphene-$h$-BN. At the end, a summary and
outlooks are given.

\section{Methods}

\subsection{The linearized quantum master equation under finite ${\bf E}$ and
the computations of $\tau_{s}$ and $l_{s}$}

\subsubsection{The linearized master equation}

We obtain the linearized master equation following the strategy of
our prior paper\citep{xu2025predicting} with a few electric-field-related
modifications. To simulate the electron transport, we apply the Wigner
transformation to the quantum (density-matrix) master equation and
obtain\citep{sekine2017quantum,xu2025predicting}
\begin{align}
 & \frac{d\rho_{\kappa}^{\mathrm{tot}}\left(t,{\bf R}\right)}{dt}+\frac{1}{2}\left\{ {\bf v}\cdot\frac{d\rho^{\mathrm{tot}}\left(t,{\bf R}\right)}{d{\bf R}}\right\} _{\kappa}\nonumber \\
= & -\frac{i}{\hbar}\left[H^{e}\left({\bf B}\right),\rho^{\mathrm{tot}}\right]_{\kappa}+D^{E}\left[\rho^{\mathrm{tot}}\right]+C_{\kappa}\left[\rho^{\mathrm{tot}}\right],\label{eq:nonlinear}
\end{align}
Where $\kappa=\left\{ k,a,b\right\} $ is the combined index of k-point
and two band indices. ${\bf R}$ is real-space coordinate. $\rho^{\mathrm{tot}}(t,{\bf R})$
is the total electronic Wigner distribution function corresponding
to the total density matrix. $\frac{1}{2}\left\{ {\bf v}\cdot\nabla_{{\bf R}}\rho^{\mathrm{tot}}\right\} _{\kappa}$
is the diffusion term with ${\bf v}$ the vector of velocity matrices.
$\left\{ {\bf a}\cdot{\bf b}\right\} $$=$${\bf a}{\bf b}$+${\bf b}{\bf a}$.
The right-hand-side terms are the coherent, drift and scattering terms
respectively. $H^{e}({\bf B})$$=H^{e}(0)$$+$$H^{\mathrm{sZ}}({\bf B})$,
with $H^{\mathrm{sZ}}({\bf B})$=$\mu_{B}g_{0}{\bf B}\cdot{\bf s}$
being spin Zeeman Hamiltonian. The drift term $D^{E}\left[\rho^{\mathrm{tot}}\right]$
under finite ${\bf E}$ along a periodic direction reads\citep{sekine2017quantum,xu2024ab,ventura2017gauge}
\begin{align}
D^{E}\left[\rho^{\mathrm{tot}}\right]= & \frac{e}{\hbar}{\bf E}\cdot\frac{D\rho^{\mathrm{tot}}}{D{\bf k}},\label{eq:DE}\\
\frac{DA_{k}}{D{\bf k}}= & \frac{dA_{k}}{d{\bf k}}-i\left[\xi_{k},A_{k}\right],\label{eq:cov-der}
\end{align}
with $\frac{DA_{k}}{D{\bf k}}$ the covariant derivative of matrix
$A_{k}$ and $\xi$ the Berry connection in the eigenstate representation.

Within Born-Markov approximation and neglecting the renormalization
part\citep{rosati2014derivation,xu2023ab,iotti2005quantum}, the scattering
term $C\left[\rho^{\mathrm{tot}}\right]$ is quadratic functional
of $\rho^{\mathrm{tot}}$ and the detailed form is given in Ref. \citenum{xu2023ab}.
The above assumptions for $C\left[\rho^{\mathrm{tot}}\right]$ are
commonly employed and typically valid in studies of $\tau_{s}$ or
$l_{s}$, which focus on slow decay processes in systems that are
usually not far from equilibrium. Their justifications were discussed
in Ref. \citenum{xu2023ab}.

Suppose $\rho^{\mathrm{tot}}$=$\rho^{\mathrm{eq}}$+$\rho$, with
$\rho^{\mathrm{eq}}$ the equilibrium density matrix. At ${\bf E}$=0,
$\rho^{\mathrm{eq}}$ becomes Fermi-Dirac function $f$. Assuming
$\rho$ is small, typical in device applications and measurements
of $\tau$ and $l$, we obtain a linearized master equation by linearizing
the scattering term\citep{xu2025predicting}
\begin{align}
\frac{d\rho_{\kappa}}{dt}+\sum_{j\kappa'}L_{\kappa\kappa'}^{v_{j}}\frac{d\rho_{\kappa'}}{dR_{j}}= & \sum_{\kappa'}L_{\kappa\kappa'}({\bf E},{\bf B})\rho_{\kappa'},\label{eq:linear}\\
L({\bf E},{\bf B})= & L^{e}({\bf B})+L^{E}({\bf E})+L^{C},
\end{align}
where $L^{v_{j}}$, $L^{e}$, $L^{E}$ and $L^{C}$ correspond to
the diffusion, coherent, drift and scattering terms respectively.
$L^{v_{j}}$ and $L^{e}$ are given in Appendix A. For the electron-phonon
(e-ph) scatting, $L^{C}$ reads
\begin{align}
L_{kab,k'cd}^{C}= & \overline{L}_{kab,k'cd}^{C}+\overline{L}_{kba,k'dc}^{C,*},\label{eq:LC}\\
\overline{L}_{kab,k'cd}^{C}= & \frac{1}{2N_{k}}\sum_{e}\left[\left(I-\rho^{\mathrm{eq}}\right)_{kae}P_{keb,k'cd}+P_{k'cd,kae}^{*}\rho_{keb}^{\mathrm{eq}}\right]\nonumber \\
 & -\frac{1}{2N_{k}}\delta_{kk'}\delta_{ac}\sum_{k''fg}P_{kdb,k''fg}\rho_{k''fg}^{\mathrm{eq}}\nonumber \\
 & -\frac{1}{2N_{k}}\delta_{kk'}\sum_{k''fg}\left(I-\rho^{\mathrm{eq}}\right)_{k''fg}P_{k''fg,kac}^{*}\delta_{bd}.\label{eq:LCbar}
\end{align}
where $P$ is the generalized scattering-rate matrix. $P$ is computed
from first-principles electron and phonon energies and the e-ph matrix
elements.\citep{xu2023ab} Eqs. \ref{eq:LC}-\ref{eq:LCbar} are slightly
different from those in our prior paper\citep{xu2025predicting},
where $\rho^{\mathrm{eq}}$ is fixed as Fermi-Dirac function $f$.
We emphasized that to obtain the linearized master equation (Eq. \ref{eq:linear})
from the nonlinear equation (Eq. \ref{eq:nonlinear}), only the scattering
term $C\left[\rho^{\mathrm{tot}}\right]$ is linearized. The other
terms in Eq. \ref{eq:nonlinear} remain unchanged but are reformulated.
Consequently, our treatment of the electric field is applicable over
a range potentially several orders of magnitude wider than that studied
here.

The quantum master equation presented above is closely connected to
standard Boltzmann transport equation and kinetic spin Bloch equation.
We compare them in Appendix B.

\subsubsection{The finite-difference computation of $L^{E}$ based on Wannier functions\label{subsec:LE}}

Given that $L^{E}\rho=D^{E}\left[\rho\right]$ and the form of $D^{E}$
in Eq. \ref{eq:DE}, to express matrix $L^{E}$, we first need to
express the covariant derivative $\frac{D\rho}{D{\bf k}}$ via finite
differences. However, a direct computation of $\frac{D\rho}{D{\bf k}}$
via Eq. \ref{eq:cov-der} is non-trivial due to the following issues:
First, the basis functions $u$ are usually obtained by diagonalizing
$H^{0}$ at different ${\bf k}$ independently, so that $u$ contain
arbitrary phase factors and are arbitrary in degenerate subspaces.
Therefore, $u$ are in general not smooth over ${\bf k}$, which makes
$\frac{d\rho}{d{\bf k}}$ not well-defined. Second, the computation
of $\xi$ usually relies on the non-degenerate perturbation theory,
which may be problematic in the presence of band degeneracies or crossings.
These issues are bypassed by using a Wannier-representation finite-difference
technique, which has been employed to compute $\frac{D\rho}{D{\bf k}}$
for nonlinear photocurrent simulation\citep{xu2024ab,silva2019high}.

To make this work self-contained, we present the expression of $\frac{D\rho}{D{\bf k}}$
follows the algorithm of our prior paper\citep{xu2024ab} below, but
with a differences: Previous work\citep{xu2024ab} relies on relaxation-time
approximation, so that all $k$-points are decoupled and $\frac{D\rho}{D{\bf k}}$
can be evaluated using finite differences between arbitrarily-close
neighboring $k$-points. Whereas here, we need to carry out finite
differences between selected $k$-points from uniform $k$ meshes
and cares must be taken for boundary $k$-points of selected ones
(see $k$-point selection in Appendix I). Additionally, we do not
directly compute $\frac{D\rho}{D{\bf k}}$ here, and instead we compute
$L^{E}$ matrix, whose formula is derived from that of $\frac{D\rho}{D{\bf k}}$.

The Wannier functions are noted as $\left|\mathbb{R}a\right\rangle $,
where $a$ is the index of a Wannier function in the unitcell and
$\mathbb{R}$ labels the unitcell. The smooth Bloch-like functions
are given by the phased sum of Wannier functions\citep{marzari2012maximally}
\begin{align}
\left|u_{ka}^{W}\right\rangle = & \sum_{\mathbb{R}}e^{-i{\bf k}\cdot\left(\widehat{{\bf r}}-\mathbb{R}\right)}\left|\mathbb{R}a\right\rangle ,\label{eq:uW}
\end{align}
which span the actual Bloch eigenstates $\left|u_{ka}\right\rangle $
at each ${\bf k}$. $\widehat{{\bf r}}$ is position operator. Here
a hat is used to emphasize that it is an operator instead of a coordinate
of electron position.

Define
\begin{align}
\widehat{H_{k}^{e}}= & e^{-i{\bf k}\cdot\widehat{{\bf r}}}\widehat{H^{e}}e^{i{\bf k}\cdot\widehat{{\bf r}}},
\end{align}
with $\widehat{H^{e}}$ the electronic Hamiltonian operator.

It follows that, if we construct the Hamiltonian in the Wannier representation
\begin{align}
H_{kab}^{W}= & \left\langle u_{ka}^{W}\right|\widehat{H_{k}^{0}}\left|u_{kb}^{W}\right\rangle ,
\end{align}
and diagonalize it as
\begin{align}
U_{k}^{\dagger}H_{k}^{W}U_{k}= & \epsilon_{k},
\end{align}
we obtain the diagonal matrix of eigenvalues $\epsilon_{k}$ and the
eigenstate matrix $U_{k}$. 

Define $\rho_{k}^{W}$=$U_{k}\rho_{k}U_{k}^{\dagger}$ as the Wigner
function in Wannier representation. According to Refs. \citenum{xu2024ab}
and \citenum{silva2019high}, the covariant derivative $\frac{D\rho}{D{\bf k}}$
satisfies:
\begin{align}
\frac{D\rho_{k}}{D{\bf k}}= & U_{k}^{\dagger}\frac{D\rho_{k}^{W}}{D{\bf k}}U_{k},\label{eq:DrhoDk_from_DrhoWDk}\\
\frac{D\rho_{k}^{W}}{D{\bf k}}= & \frac{d\rho_{k}^{W}}{d{\bf k}}-i\left[\xi_{k}^{W},\rho_{k}^{W}\right],
\end{align}
where $\xi^{W}$ is Berry connection in Wannier representation. In
Wannier representation, the basis $u_{ka}^{W}$ are smooth over ${\bf k}$,
so that $\frac{D\rho_{k}^{W}}{D{\bf k}}$ can be accurately computed
via finite differences:
\begin{align}
\frac{D\rho_{k}^{W}}{Dk_{\alpha}}= & \frac{\rho_{k_{\alpha}^{+}}^{W}-\rho_{k_{\alpha}^{-}}^{W}}{2dk_{\alpha}}-i\left[\xi_{k\alpha}^{W},\rho_{k}^{W}\right],\label{eq:DrhoWDk}
\end{align}
where $k_{\alpha}^{\pm}=k_{\alpha}\pm dk_{\alpha}$.

From Eqs. \ref{eq:DE}, \ref{eq:DrhoDk_from_DrhoWDk} and \ref{eq:DrhoWDk},
we have
\begin{align}
D^{E}\left[\rho\right]= & \sum_{\alpha}\frac{eE_{\alpha}}{\hbar}\frac{o^{kk_{\alpha}^{+}}\rho_{k_{\alpha}^{+}}o^{kk_{\alpha}^{+},\dagger}-o^{kk_{\alpha}^{-}}\rho_{k_{\alpha}^{-}}o^{kk_{\alpha}^{-},\dagger}}{2d{\bf k}}\nonumber \\
 & -i\sum_{\alpha}\frac{eE_{\alpha}}{\hbar}\left[\overline{\xi}_{k\alpha},\rho_{k}\right],\\
o^{k_{1}k_{2}}= & U_{k_{1}}^{\dagger}U_{k_{2}},\\
\overline{\xi}_{k}= & U_{k}^{\dagger}\xi_{k}^{W}U_{k}.
\end{align}

Therefore,
\begin{align}
L^{E}= & \sum_{\alpha}E_{\alpha}L_{\alpha}^{E},\\
L_{\alpha,kabk'cd}^{E}= & \frac{e}{\hbar}\frac{o_{ac}^{kk_{\alpha}^{+}}\left(o^{kk_{\alpha}^{+},\dagger}\right)_{db}}{2d{\bf k}}\delta_{k'k_{\alpha}^{+}}\nonumber \\
 & -\frac{e}{\hbar}\frac{o_{ac}^{kk_{\alpha}^{-}}\left(o^{kk_{\alpha}^{-},\dagger}\right)_{db}}{2d{\bf k}}\delta_{k'k_{\alpha}^{-}}\nonumber \\
 & -\frac{ie}{\hbar}\left(\overline{\xi}_{kac}\delta_{bd}-\delta_{ac}\overline{\xi}_{kdb}\right)\delta_{kk'}.
\end{align}

For boundary $k$-points, the derivatives are expressed via simple
finite differences instead of central ones, so that the formula of
$L^{E}$ is slightly different from the above formula but its derivation
is straightforward.

\subsubsection{The solution of $\rho^{\mathrm{eq}}$ under finite ${\bf E}$\label{subsec:solve_rhoeq_E}}

Since $\rho^{\mathrm{eq}}$ is here independent of both $t$ and ${\bf R}$,
from Eq. \ref{eq:nonlinear}, $\rho^{\mathrm{eq}}$ satisfies
\begin{align}
-\frac{i}{\hbar}\left[H^{e},\rho^{\mathrm{eq}}\right]+D^{E}\left[\rho^{\mathrm{eq}}\right]+C\left[\rho^{\mathrm{eq}}\right]= & 0.\label{eq:rhoeq1}
\end{align}

Suppose $\rho^{\mathrm{eq}}$=$f$+$\delta\rho^{\mathrm{eq}}$. $C\left[\rho^{\mathrm{eq}}\right]$
can be separated into a linear part $L^{C}\left[f\right]$ and a quadratic
part $M\left[\delta\rho^{\mathrm{eq}}\right]$,
\begin{align}
C\left[\rho^{\mathrm{eq}}\right]= & L^{C}\left[f\right]\delta\rho^{\mathrm{eq}}+M\left[\delta\rho^{\mathrm{eq}}\right].\label{eq:C_as_LC_M}
\end{align}
\begin{align}
M_{kab}\left[\delta\rho^{\mathrm{eq}}\right]= & \frac{1}{2}\sum_{ck'de}\left(\begin{array}{c}
-\rho_{kac}P_{kcb,k'de}\rho_{k'de}\\
+\rho_{k'de}P_{k'de,kac}^{*}\rho_{kcb}
\end{array}\right)+H.C..
\end{align}

Note that Eq. \ref{eq:C_as_LC_M} is exact and is applicable to large
$\delta\rho^{\mathrm{eq}}$. In Eq. \ref{eq:C_as_LC_M}, the coefficient
matrix of the linear part - $L^{C}\left[f\right]$ is a fixed matrix,
which facilitates the iterative solution for $\rho^{\mathrm{eq}}$
described in the following.

To derive the above equations, we have considered $C\left[f\right]$=0,
reflecting the fact that there is no scattering at equilibrium without
electric field.

Considering that $D^{E}\left[\rho^{\mathrm{eq}}\right]$=$L^{E}f$+$L^{E}\delta\rho^{\mathrm{eq}}$
and defining $\tilde{L}\delta\rho^{\mathrm{eq}}$=$(L^{e}+L^{E}+L^{C}\left[f\right])\delta\rho^{\mathrm{eq}}$,
from Eq. \ref{eq:rhoeq1}, we have
\begin{align}
L^{E}f+\tilde{L}\delta\rho^{\mathrm{eq}}+M\left[\delta\rho^{\mathrm{eq}}\right]= & 0.\label{eq:rhoeq2}
\end{align}

This can be rewritten as
\begin{align}
\delta\rho^{\mathrm{eq}}= & -(\tilde{L})^{-1}\left\{ L^{E}f+M\left[\delta\rho^{\mathrm{eq}}\right]\right\} .
\end{align}

Therefore, $\delta\rho^{\mathrm{eq}}$ can be solved by fixed-point
iteration with a mixing parameter $\alpha^{\mathrm{mix}}$:
\begin{align}
\delta\rho^{\mathrm{eq},(n+1)}= & \left(1-\alpha^{\mathrm{mix}}\right)\delta\rho^{\mathrm{eq},(n)}+\alpha^{\mathrm{mix}}\delta\rho_{0}^{\mathrm{eq},(n+1)},
\end{align}
\begin{align}
\delta\rho_{0}^{\mathrm{eq},(n+1)}= & -(\tilde{L})^{-1}\left\{ L^{E}f+M\left[\delta\rho^{\mathrm{eq},(n)}\right]\right\} ,\\
\delta\rho_{0}^{\mathrm{eq},(1)}= & -(\tilde{L})^{-1}L^{E}f,
\end{align}
with $\delta\rho^{\mathrm{eq},(n)}$ being $\delta\rho^{\mathrm{eq}}$
at $n$th iteration. In all calculations with ${\bf E}$$\neq$0,
we numerically iterate until the absolute and relative errors of $\delta\rho^{\mathrm{eq}}$
fall below 10$^{-12}$ and 10$^{-6}$ respectively. The fully converged
$\delta\rho^{\mathrm{eq}}$ is then used to compute the scattering
term $L^{C}\left[\rho^{\mathrm{eq}}\right]$ (the other terms in Eq.
\ref{eq:linear} do not depend on $\rho^{\mathrm{eq}}$). According
to our numerical tests, we find that $-(\tilde{L})^{-1}L^{E}f$ is
already a good approximation of $\delta\rho^{\mathrm{eq}}$ within
our studied electric-field range.

\subsubsection{The computations of $\tau_{s}$ and $l_{s}$ via solving a full EVP\label{subsec:full}}

This part was described in our prior paper\citep{xu2025predicting},
we present it here to make this work self-contained. To obtain $\tau_{s}$
and $l_{s}$, it is required to solve linearized master equation Eq.
\ref{eq:linear}. Here we consider two important commonly used one-variable
problems: (i) spatial homogeneous relaxation and (ii) 1D steady-state
diffusion along $x$ direction. The solutions for the relaxation and
diffusion problems are\citep{xu2025predicting}
\begin{align}
\rho_{\kappa}(t)= & \Sigma_{\mu}e^{-\Gamma_{\mu}t}U_{\kappa\mu}^{tR}c_{\mu}^{t},\label{eq:general_solution}\\
\rho_{\kappa}(x)= & \Sigma_{\nu}e^{-\lambda_{\nu}^{x}x}U_{\kappa\nu}^{xR}c_{\nu}^{x},\label{eq:general_solution_2}
\end{align}
respectively. $\mu$ and $\nu$ are decay mode indices. $\Gamma_{\mu}$
is complex relaxation rate and $\lambda_{\mu}^{x}$ is called complex
diffusion rate here. $\mathrm{Re}\mathrm{\Gamma}_{\mu}$ and $\mathrm{Re}\lambda_{\mu}^{x}$
are decay-mode-resolved lifetime $\tau_{\mu}$ and inverse diffusion
length $1/l_{\nu}^{x}$, respectively. $\mathrm{Im}\Gamma_{\mu}$
and $\mathrm{Im}\lambda_{\nu}^{x}$ describe temporal and spatial
precession, respectively. $U_{\kappa\mu}^{tR}$ are right eigenvectors
of standard EVP with $\Gamma_{\mu}$ being the eigenvalues:
\begin{align}
-\Sigma_{\kappa'}L_{\kappa\kappa'}U_{\kappa'\mu}^{tR}= & U_{\kappa\mu}^{tR}\Gamma_{\mu}.\label{eq:sevp}
\end{align}
and $U_{\kappa\nu}^{xR}$ are right eigenvectors of the generalized
EVP with $E_{\nu}^{X}$=1/$\lambda_{\nu}^{x}$ being the eigenvalues:
\begin{align}
\Sigma_{\kappa'}L_{\kappa\kappa'}^{v_{x}}U_{\kappa'\nu}^{xR}= & -\Sigma_{\kappa'}L_{\kappa\kappa'}U_{\kappa'\nu}^{xR}(\lambda_{\nu}^{x})^{-1}.\label{eq:gevp}
\end{align}

We consider the boundary conditions: $\rho$($t$=0$)$=$\rho^{s_{i}}$
and $\rho$($t$$\rightarrow$$\infty$$)$=0 for relaxation problem;
$\rho$($x$=0$)$=$\rho^{s_{i}}$ and $\rho$($x$$\rightarrow$$\infty$$)$=0
for diffusion problem. $\rho^{s_{i}}$ (Eq. \ref{eq:rhos}) is spin
perturbative density-matrix induced by spin Zeeman effect\citep{xu2020spin}.
We then obtain $c_{\mu}^{t}$=$-\sum_{\kappa}U_{\kappa\mu}^{tL,*}\rho_{\kappa}^{\mathrm{pert}}$
with $U_{\kappa\mu}^{tL}$ left eigenvector of Eq. \ref{eq:sevp},
and $c_{\nu}^{x}$=$-\sum_{\kappa\kappa'}U_{\kappa\nu}^{xL,*}L_{\kappa\kappa'}\rho_{\kappa'}^{\mathrm{pert}}$
with $U_{\kappa\nu}^{xL}$ left eigenvector of Eq. \ref{eq:gevp}.
With the solutions of $\rho_{\kappa}(t)$ or $\rho_{\kappa}(x)$,
we can then simulate the decay curves of spin observable and obtain
$\tau_{s}$ and $l_{s}$. See more details in Appendix D.

\subsection{Approximate methods of $\tau_{s}\left(E\right)$ and $l_{s}\left(E\right)$}

\subsubsection{Approximate calculation based on Rayleigh-Ritz (RR) method\label{subsec:RR}}

Low-power electronics often require slow decay of quantities like
spin for stable detection and manipulation of information. For slow
decay, it seems unnecessary to solve full EVPs (Eq. \ref{eq:gevp}),
but enough to obtain eigenvalues and eigenvectors of a few ``relevant''
modes using approximate methods, e.g., the RR method\citep{macdonald1933successive}.
In this method, eigenvectors are linear combinations of trial vectors,
which span a Krylov subspace (KS)\citep{liesen2013krylov} here. For
a full EVP $AU^{R}=BU^{R}E$ (Eq. \ref{eq:gevp}), order-$n$ right
KS $V^{KR}$ consists of $A^{m-l}B^{l}V^{R}$ with $0\le l\le m\le n$,
where columns of $V^{R}$ are trial vectors. Similarly, left KS $V^{KL}$
consists of $A^{\dagger,m-l}B^{\dagger,l}V^{L}$.

With $V^{KR\left(L\right)}$ and $M^{K}=V^{KL,\dagger}MV^{KR}$, a
reduced EVP is obtained
\begin{align}
A^{K}Y^{R}= & B^{K}Y^{R}E.\label{eq:reduced_gep}
\end{align}

Eigenvalues of Eq. \ref{eq:reduced_gep} are approximate eigenvalues
of the full EVP (Eq. \ref{eq:gevp}). Eigenvectors $U^{R(L)}$$\approx$$V^{KR(L)}Y^{R(L)}$
if enforcing $V^{KL,\dagger}B^{K}V^{KR}$=$I$.

Below we specify a few types of trial vectors $V^{R(L)}$ for approximate
spin decay simulations based on low-order RR method.

\textbf{Spin Relaxation:}

The corresponding reduced EVP is
\begin{align}
-\left(L^{E0K}+L^{EK}\right)Y^{R}= & I^{K}Y^{R}\Gamma_{s},\label{eq:RR_EVP_taus}
\end{align}
where $L^{E0K}$=$V^{KL,\dagger}L^{E0}V^{KR}$ with $L^{E0}$ is $L$
matrix at ${\bf E}$=0, $L^{EK}$=$V^{KL,\dagger}L^{E}V^{KR}$ and
$I^{K}$=$V^{KL,\dagger}V^{KR}$. Here we set
\begin{align}
V^{R(L)}= & \left\{ U_{sx}^{tR(L)},U_{sy}^{tR(L)},U_{sz}^{tR(L)}\right\} ,\label{eq:RR_taus}
\end{align}
where $U_{s\alpha}^{tR(L)}$ is the right (left) eigenvector of Eq.
\ref{eq:sevp} for zero-field relaxation problem, with the eigenvalue
being zero-electric-field spin relaxation rate of $S_{\alpha}$ -
$\Gamma_{s,\alpha}^{0}$. We refer to simulations using order-$n$
RR method with such $V^{R(L)}$ as ``$n$-RR'' simulations here.

\textbf{Spin Diffusion:}

The corresponding reduced EVP is
\begin{align}
-\left(L^{E0K}+L^{EK}\right)Y^{R}= & L^{v_{x}K}Y^{R}\lambda_{s},\label{eq:RR_EVP_ls}
\end{align}
where $L^{v_{x}K}$=$V^{KL,\dagger}L^{v_{x}}V^{KR}$. To emphasize
the importance of $V^{R(L)}$ to the RR-based simulation of $l_{s}$,
we consider three distinct types of $V^{R(L)}$:

(i) $n$-RR simulation.
\begin{align}
V^{R(L)}= & \left\{ U_{sx\pm}^{xR(L)},U_{sy\pm}^{xR(L)},U_{sz\pm}^{xR(L)}\right\} ,\label{eq:RR_ls}
\end{align}
where $U_{s\alpha\pm}^{xR(L)}$ is the right (left) eigenvector of
Eq. \ref{eq:sevp} for zero-field diffusion problem, with the eigenvalue
being $\pm1/\lambda_{s,\alpha}^{0}$. $\lambda_{s,\alpha}^{0}$ is
zero-electric-field (complex) spin diffusion rate of $S_{\alpha}$.

(ii) $n$-RR-B simulation.
\begin{align}
V^{R(L)}= & \left\{ U_{s}^{tR(L)},L^{v_{x}}U_{s}^{tR(L)}\right\} ,\label{eq:RR-B}\\
U_{s}^{tR(L)}= & \left\{ U_{sx}^{xR(L)},U_{sy}^{xR(L)},U_{sz}^{xR(L)}\right\} .
\end{align}

(iii) $n$-RR-C simulation.
\begin{align}
V^{R(L)}= & \left\{ U_{s}^{cR(L)},L^{v_{x}}U_{s}^{cR(L)}\right\} ,\label{eq:RR-C}\\
U_{s}^{cR}= & \left\{ \rho^{s_{x}},\rho^{s_{y}},\rho^{s_{z}}\right\} ,\\
U_{s}^{cL}= & \left\{ \varrho^{s_{x}},\varrho^{s_{y}},\varrho^{s_{z}}\right\} ,\label{eq:RR-C3}
\end{align}
where $\rho^{s_{\alpha}}$ (Eq. \ref{eq:rhos}) is spin perturbative
density matrix introduced above in Sec. \ref{subsec:full}. Its dual
vector $\varrho_{s}^{s_{\alpha}}$ satisfies $\varrho_{s}^{s_{\alpha}}$$\propto$$s_{\alpha}$
and $\left\langle \varrho_{s}^{s_{\alpha}}|\rho^{s_{\alpha}}\right\rangle $=1.
According to our prior paper (Ref. \citenum{xu2025predicting}), $\rho^{s_{\alpha}}$
and $\varrho_{s}^{s_{\alpha}}$ are approximate right and left eigenvectors
of the scattering-term matrix $L^{C}$, with the corresponding eigenvalue
$\Gamma_{\alpha}^{\mathrm{EY}}$ approximating the Elliott-Yafet (EY)-type
spin relaxation rate arising from spin-flip scattering.

We emphasize that the RR method with KS is a powerful tool for estimating
$\tau_{s}$ and $l_{s}$. Low-order RR simulations are valuable to
provide mechanistic insights, while high-order RR simulations generally
improve accuracy. However, the method has limitations: (i) Its accuracy
strongly depends on the choices of both $V^{R}$ and $V^{L}$. This
is particularly critical for low-order simulations. (ii) There is
no general guarantee that $V^{KR}$ and $V^{KL}$ simultaneously form
suitable basis sets for the true right and left eigenvectors, which
may lead to significant errors of $\Gamma_{s}$, $\lambda_{s}$ and
the corresponding eigenvectors. This issue may be mitigated by employing
specific techniques in the KS-based (block) iterative eigensolvers,
such as updating the basis according to the residual of the approximate
EVP.

\subsubsection{The \textit{ab initio} matrix-drift-diffusion (ab-mDD) model\label{subsec:ab-mDD}}

Here we derive an advanced theoretical model from linearized master
equation. It generalizes the standard drift-diffusion model with several
additional physical effects.

Suppose $V_{\kappa\nu}^{sR}$ and $V_{\kappa\nu}^{sL}$ ($\nu$=1,2,...,$N^{s}$)
are biorthonormal basis functions of $\rho$ ($\left\langle V_{\kappa\nu}^{sL}|V_{\kappa\nu'}^{sR}\right\rangle $=$\delta_{\nu\nu'}$)
and are highly relevant to spin decay. Define a projector $P$ and
its complement $Q$ as
\begin{align}
P_{\kappa\kappa'}= & \sum_{\nu}\left|V_{\kappa\nu}^{sR}\right\rangle \left\langle V_{\kappa'\nu}^{sL}\right|\\
Q= & I-P,
\end{align}
so that any matrix $M$ can be separated into $M^{P}$+$M^{Q}$ with
$M^{P}=PM$ and $M^{Q}=QM$.

To derive an approximate master equation for spin decay, we introduce
the following ansatz -

\textbf{Ansatz I}: The steady-state Wigner function $\rho$ consists
of a dominant part $\rho^{P}$=$P\rho$ and a small correction $\delta$=$Q\rho$,
i.e.,
\begin{align}
\rho\left(t,{\bf R}\right)= & \rho^{P}\left(t,{\bf R}\right)+\delta\left(t,{\bf R}\right),\label{eq:rho_abmDD}
\end{align}
with $|\delta|\ll|\rho^{P}|$. Define $S_{\nu}\left(t,{\bf R}\right)=\left\langle V_{\kappa\nu}^{sL}|\rho_{\kappa}\right\rangle $,
then spin evolution can be well described by $S_{\nu}\left(t,{\bf R}\right)$.

In this work, we adopt a specific choice:

\textbf{Ansatz II}: $V_{\kappa\alpha}^{sR}=\rho^{s_{\alpha}}$ and
$V_{\kappa\alpha}^{sL}=\varrho^{s_{\alpha}}$ with $\alpha$=$x,y,z$
(see the text below Eq. \ref{eq:RR-C3} in Sec. \ref{subsec:RR}).

Consequently, the linearized master equation, Eq. \ref{eq:linear},
can be written as (noting that $L^{C,Q}\rho^{P}=0$ and $L^{C,P}\delta=0$)
\begin{align}
\left\{ \begin{array}{c}
\frac{d\rho^{P}}{dt}+\frac{d\delta}{dt}+\\
\left[L^{{\bf v}}\cdot\nabla_{{\bf R}}-\left(L^{e,Q}+L^{E,Q}\right)\right]\rho^{P}\\
+\left[L^{{\bf v}}\cdot\nabla_{{\bf R}}-\left(L^{e}+L^{E}\right)\right]\delta
\end{array}\right\} = & L^{P}\rho^{P}+L^{C,Q}\delta.\label{eq:proj_ldmme}
\end{align}

To proceed to a reduced master equation of only $S_{\alpha}\left(t,{\bf R}\right)$,
we need to consider relative magnitudes of different terms in the
above equation and make the following ansatz -

\textbf{Ansatz III}: \textbf{$|d\rho^{P}/dt|$, $|L^{P}\rho^{P}|$
and the left 4th term are much smaller than $|L^{C,Q}\delta|$ and
the left 3rd term.} This condition is typically satisfied in practice;
the justification is provided in Appendix E.

Based on these three ansatzs, we obtain a reduced master equation
of $S_{\alpha}({\bf R})$ (see derivation in Appendix F):
\begin{align}
\frac{dS_{\alpha}}{dt}= & D_{ij,\alpha\beta}\frac{d^{2}S_{\beta}}{dR_{i}dR_{j}}+v_{j,\alpha\beta}^{d}\left({\bf E}\right)\frac{dS_{\beta}}{dR_{j}}-\Gamma_{\alpha\beta}^{P}\left({\bf E}\right)S_{\beta},\label{eq:ab-mDD}
\end{align}
where $D_{ij}$, $v_{j}^{d}\left({\bf E}\right)$ and $\Gamma^{P}\left({\bf E}\right)$
are effective diffusion-coefficient, drift-velocity and scattering-rate
matrices (with indices corresponding to spin directions). Define $T=-(L^{C,Q})^{-1}$,
we have
\begin{align}
D_{ij}= & \varrho^{s,\dagger}L^{v_{i}}TL^{v_{j}}\rho^{s},\label{eq:Dij}\\
v_{j}^{d}\left({\bf E}\right)= & E_{j}\widetilde{\mu}_{j}+v_{j}^{d0},\label{eq:vdj}\\
\widetilde{\mu}_{j}= & -E_{j}^{-1}\varrho^{s,\dagger}\left(L^{v_{j}}TL^{E,Q}+L^{E}TL^{v_{j}}\right)\rho^{s},\label{eq:muj_tilde}\\
v_{j}^{d0}= & -\varrho^{s,\dagger}\left(L^{v_{j}}TL^{e,Q}+L^{e}TL^{v_{j}}\right)\rho^{s},\label{eq:vd0}\\
\Gamma^{P}\left({\bf E}\right)= & \Gamma^{0}+\Omega^{t}\left({\bf E}\right)+\Gamma^{PE}\left({\bf E}\right),\label{eq:GP}\\
\Gamma^{\mathrm{0}}= & \varrho^{s,\dagger}\left(-L^{C,P}-L^{e}TL^{e,Q}\right)\rho^{s},\\
\Omega^{t}\left({\bf E}\right)= & \varrho^{s,\dagger}\left(L^{e,P}+L^{E,P}\right)\rho^{s}\label{eq:Omegat}\\
 & -\varrho^{s,\dagger}\left(L^{E}TL^{e,Q}+L^{e}TL^{E,Q}\right)\rho^{s},\\
\Gamma^{E}\left({\bf E}\right)= & -\varrho^{s,\dagger}L^{E}TL^{E,Q}\rho^{s}.\label{eq:GE}
\end{align}

The physical interpretations of these terms are as follows:

(i) Practically, $T$ can be well approximated as the momentum lifetime
$\tau_{m}$, so that the effective diffusion coefficient $D_{jj\alpha\alpha}$$\approx$$v_{Fjj}^{2}\tau_{m}$
with $v_{Fjj}^{2}$ being $R_{j}$-component of Fermi velocity square.
In this work, $T$ is computed by inverting $L^{C}$ matrix, although
using $T$$\approx$$\tau_{m}$ only slightly alters the final results.

(ii) $\widetilde{\mu}_{j}$ is effective mobility and can be regarded
as carrier mobility or its negative values for conduction electrons
or holes respectively in semiconductors. According to Appendix G,
we approximately have $\widetilde{\mu}_{j}$$\approx$$e\tau_{m}/\widetilde{m}_{j}$,
with $\widetilde{m}_{j}^{-1}$=$\varrho_{s}^{\dagger}(\hbar^{-1}dv_{kj}/\hbar dk_{j})\rho_{s}$.

(iii) $v_{j}^{d0}$ can be expressed as $D_{jj}\Omega^{R_{j}}$, where
$\Omega^{R_{j}}$ accounts for zero-magnetic-field spatial spin precession
due to spin-orbit fields.

(iv) $\Gamma^{\mathrm{0}}$ is zero-electric-field spin relaxation
rate. $\Gamma^{0}$ is actually the sum of EY spin relaxation rate
$\Gamma^{\mathrm{EY}}$=$-\varrho^{s,\dagger}L^{C,P}\rho^{s}$ and
D\textquoteright yakonov-Perel\textquoteright{} (DP) spin relaxation
rate $\Gamma^{\mathrm{DP}}$=$-\varrho^{s,\dagger}L^{e}TL^{e,Q}\rho^{s}$$\approx$$\tau_{m}$$\Omega^{2}$
with $\Omega^{2}$=$-\varrho^{s,\dagger}L^{e}L^{e,Q}\rho^{s}$. DP
spin relaxation arises from random spin precession between adjacent
scattering events.

(v) $\Omega^{t}$ describes temporal spin precession, arising not
only from an external magnetic field (Zeeman term in $L^{e,P}$) and
magnetic moment in magnet (via $L^{e,P}$), but also from an electric-field-induced
effective magnetic field (via $L^{E}TL^{e,Q}$+$L^{e}TL^{E,Q}$).

(vi) $\Gamma^{E}$ is electric-field-induced DP spin relaxation rate.

In general, the ab-mDD model (Eq. \ref{eq:ab-mDD}) does not have
an explicit solution and needs to be solved numerically. For spatial
homogeneous relaxation and 1D steady-state diffusion, there exists
explicit solutions. See details in Appendix H.

The ab-mDD model is similar to low-order RR method (Sec. \ref{subsec:RR})
in two aspects: (i) they both approximate the system spin dynamics
using a limited number of degrees of freedom; (ii) for spin diffusion,
they involve similar matrix elements. For instance, 1-RR-C (Eqs. \ref{eq:RR-C}-\ref{eq:RR-C3})
approximately contains $\varrho^{s,\dagger}L^{e}L^{e}\rho^{s}$$\approx$$\varrho^{s,\dagger}L^{e}L^{e,Q}\rho^{s}$,
which also appears in the ab-mDD model if $T$ is approximated as
$\tau_{m}$. Numerically, however, they are not identical and may
show quite different results; the ab-mDD model seems often more accurate
than 1-RR-C. The ab-mDD model has the advantage of clear physical
interpretability, since it is a generalization of the standard drift-diffusion
model with additional key effects from the terms $v_{j}^{d0}$, $\Omega^{t}$
and $\Gamma^{PE}$. While low-RR method has the advantage that it
can be systematically improved by selecting better $V^{R(L)}$ and
increasing the order.

\begin{figure*}
\includegraphics[scale=0.6]{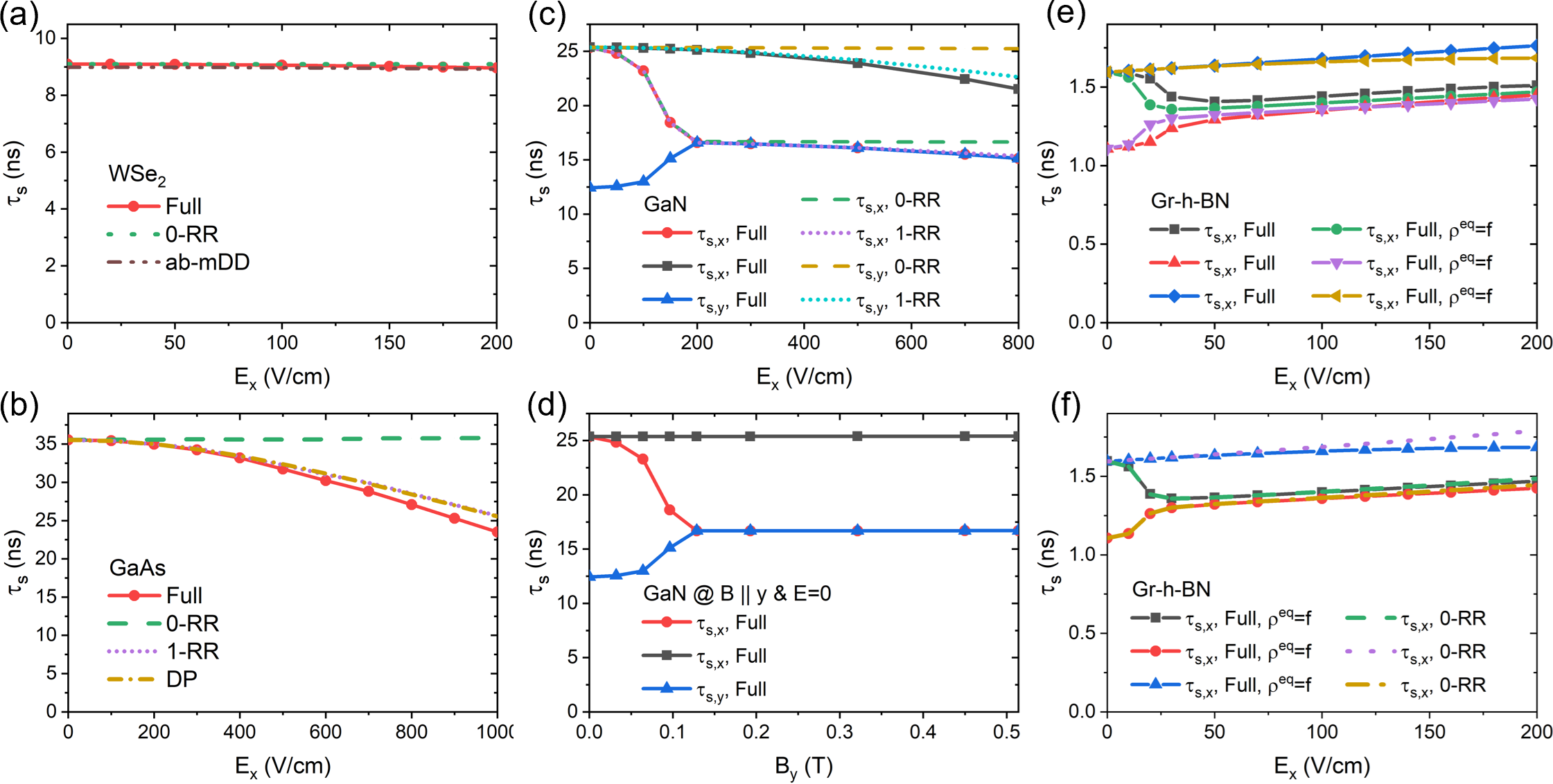}

\caption{Calculated electric-field ($E_{x}$) dependent spin lifetimes ($\tau_{s}(E_{x})$)
by different methods. $\tau_{s}(E_{x})$ of (a) holes of WSe$_{2}$
at 50 K, (b) electrons of GaAs at 300 K, and (c) electrons of GaN
at 100 K. (d) $\tau_{s}\left(E_{x}=0\right)$ of GaN at different
$B_{y}$ (external magnetic field along $y$). (e) and (f) are $\tau_{s}\left(E_{x}\right)$
of graphene-$h$-BN at 300 K with Fermi level $E_{F}$=0.1 eV. ``Full''
means that $\tau_{s}$ is calculated by the full \textit{ab initio}
approach solving full EVP Eq. \ref{eq:sevp}, with a self-consistently-computed
field-dependent equilibrium density matrix $\rho^{\mathrm{eq}}$ (see
Sec. \ref{subsec:solve_rhoeq_E}). ``$n$-RR'' means that $\tau_{s}(E_{x})$
is computed by Eq. \ref{eq:RR_EVP_ls} using $n$th-order RR method
considering a certain number of spin ``relevant'' decay modes (see
Sec. \ref{subsec:RR}). ``ab-mDD'' means our proposed \textit{ab initio}
matrix-drift-diffusion model (Sec. \ref{subsec:ab-mDD} and Eq. \ref{eq:ab-mDD}).
``DP'' corresponds to Eq. \ref{eq:tausE_DP}, which is a DP-like
model derived from 1-RR. ``Full, $\rho^{\mathrm{eq}}$=$f$'' means
that $\tau_{s}$ is calculated by solving full EVP Eq. \ref{eq:sevp},
but with $\rho^{\mathrm{eq}}$ being its zero-electric-field value
- Fermi-Dirac function $f$.\label{fig:taus}}
\end{figure*}

It is possible to derive other drift-diffusion models by decomposing
$\rho$ into more than two parts, choosing different $V^{sR(L)}$,
or reorganizing the terms of Eq. \ref{eq:ab-mDD} differently. However,
such alternatives may encounter specific issues. For example, if we
set $V_{\kappa\alpha}^{sR(L)}$ as $U_{s\alpha,\kappa}^{tR(L)}$ (eigenvector
corresponding to $\Gamma_{s,\alpha}^{0}$), the term $v_{j}^{d0}$
in the resulting drift-diffusion model will be $-U_{s}^{tL,\dagger}L^{v_{j}}TL^{e,Q}U_{s}^{tR}$,
which is approximately half of $v_{j}^{d0}$ (Eq. \ref{eq:vd0}) in
our proposed ab-mDD model and thus underestimates the significance
of zero-magnetic-field spatial spin precession.

\subsubsection{Three model formulas of $l_{s}\left(E\right)$ with simple analytic
expressions}

Eq. \ref{eq:lsE_from_standard_DD} can be rewritten as
\begin{align}
\frac{1}{l_{s}\left(E\right)}= & \sqrt{\frac{1}{l_{s0}^{2}}+\frac{E^{2}}{4}\eta^{2}}-\theta\frac{\left|E\right|}{2}\eta,\label{eq:model}
\end{align}
where the parameter $\eta$ is a positive constant and may be estimated
in different ways, leading to different results of $l_{s}\left(E\right)$.
In this work, we consider three types of $\eta$--denoted $\eta_{1}$,
$\eta_{2}$ and $\eta_{3}$:

(1) For 1D steady-state diffusion along $x$ direction, neglecting
the terms $v_{j}^{d0}$, $\Omega^{t}$ and $\Gamma^{PE}$, the ab-mDD
model (Sec. \ref{subsec:ab-mDD}) reduces to:
\begin{align}
D_{xx}\frac{d^{2}S}{dx^{2}}+\widetilde{\mu}_{x}\frac{dS}{dx}= & \frac{S}{\tau_{s0}},
\end{align}
which is the standard drift-diffusion model at zero magnetic field,
with $D_{xx}$ and $\widetilde{\mu}_{x}$ being defined by Eqs. \ref{eq:Dij}
and \ref{eq:muj_tilde}. $\tau_{s0}$ is $\tau_{s}$ at ${\bf E}$=0.
Using the expressions $D_{jj\alpha\alpha}$$\approx$$v_{Fjj}^{2}\tau_{m}$
and $\widetilde{\mu}_{j}$$\approx$$\tau_{m}/\widetilde{m}_{j}$
given below Eq. \ref{eq:GE} in Sec. \ref{subsec:ab-mDD}, we define
\begin{align}
\eta_{1}= & e/\left(\left|\widetilde{m}_{x}\right|v_{Fx}^{2}\right).\label{eq:model1}
\end{align}

For a semiconductor, $\widetilde{m}$ can be regarded as Fermi-surface-averaged
effective mass or its negative values, for conduction electrons or
holes respectively.

(2) Instead of using carrier diffusion coefficient, we use an effective
spin diffusion coefficient $D_{s}$ defined from \textit{ab initio}
$l_{s0}$ and $\tau_{s}$ - $D_{s}=l_{s0}^{2}/\tau_{s}$. $\eta$
is then
\begin{align}
\eta_{2}= & \mu/D_{s},\label{eq:model2}
\end{align}
with $\mu$ computed via first-principles linearized Boltzmann transport
equation.\citep{ponce2020first}

(3) Using Einstein relation for non-degenerate semiconductors, we
set $\eta$ as
\begin{align}
\eta_{3}= & e/\left(k_{B}T\right),\label{eq:model3}
\end{align}
which was previously used in Refs. \citenum{yu2002electric} and \citenum{zucchetti2022electric}.
The above equation can be improved by using generalized Einstein relation
- $D\approx-\mu n\frac{dn}{dE_{F}}$ for both degenerate and non-degenerate
semiconductors. However, we find that this improvement does not change
the conclusions of our study, so that we retain Eq. \ref{eq:model3}
for simplicity.

In the following, we refer to ``Model 1, 2 and 3'' as the models
which all use Eq. \ref{eq:model} but with $\eta$ being set as $\eta_{1}$,
$\eta_{2}$ and $\eta_{3}$ respectively.

\section{Results and Discussions}

In this work, we study $\tau_{s}$ and $l_{s}$ at finite $E_{x}$
(electric field along $x$ direction) for four representative systems
- holes of weakly-doped (or nondegenerate) monolayer $p$-type WSe$_{2}$,
electrons of weakly-doped bulk $n$-type GaAs and GaN, and graphene-$h$-BN
heterostructure with $E_{F}$=0.1 eV. These systems are chosen to
encompass different spin relaxation mechanisms, types of spin-orbit
fields, and band structures. Computational details are given in Appendix
I. We have not considered a metal, since the electric-field effect
for metals is generally believed to be very weak\citep{yu2002electric}
and our theoretical results indeed show that the electric-field effects
on both $\tau_{s}$ and $l_{s}$ of Pt at 300 K are negligible even
at $E_{x}$=10$^{4}$ V/cm.

We focus on spin decay driven by spin-orbit coupling and the e-ph
scattering, incorporating the drift dynamics due to electric fields.
The electron-impurity and electron-electron scatterings are not considered
for simplicity. In the main text, we present and compare results obtained
from the full \textit{ab initio} approach and various approximate
methods, aiming to extract underlying physical mechanisms The comparison
between theoretical and experimental data involves multiple complexities
and is discussed separately in Appendix\,J. The $E_{x}$ range is
chosen such that the maximum value of approximate $l_{s}$ estimated
from 0-RR simulation is about 3$l_{s0}$. The employed quantum master
equation is applicable to electric field much higher (possibly by
a few orders of magnitude) than the $E_{x}$ range here. But the applicability
of its numerical implementation is limited by the finite-difference
errors in computing $L^{E}$ (Sec. \ref{subsec:LE}). We have verified
that these errors are negligible within the $E_{x}$ range studied.

\subsection{The electric-field effect on $\tau_{s}$}

From Fig. \ref{fig:taus}(a), it is clear that the electric-field
effect on $\tau_{s}$ of WSe$_{2}$ is weak. Even at 1000 V/cm (not
shown), $\tau_{s}$ decreases by only 8\%. This is expected because
spins of WSe$_{2}$ are highly polarized and the electric-field hardly
alters the spin polarization. The slight reduction at higher electric
field of 1000 V/cm likely occurs because the drift term pushes electrons
toward larger wavevectors (relative to $\pm K$), where spin-flip
scattering is stronger. This then enhances EY spin relaxation and
reduces $\tau_{s}$. $\tau_{s}(E_{x})$ of WSe$_{2}$ is well reproduced
by both RR method and the ab-mDD model, confirming the validity of
those approximate methods for EY systems.

\begin{figure*}
\includegraphics[scale=0.6]{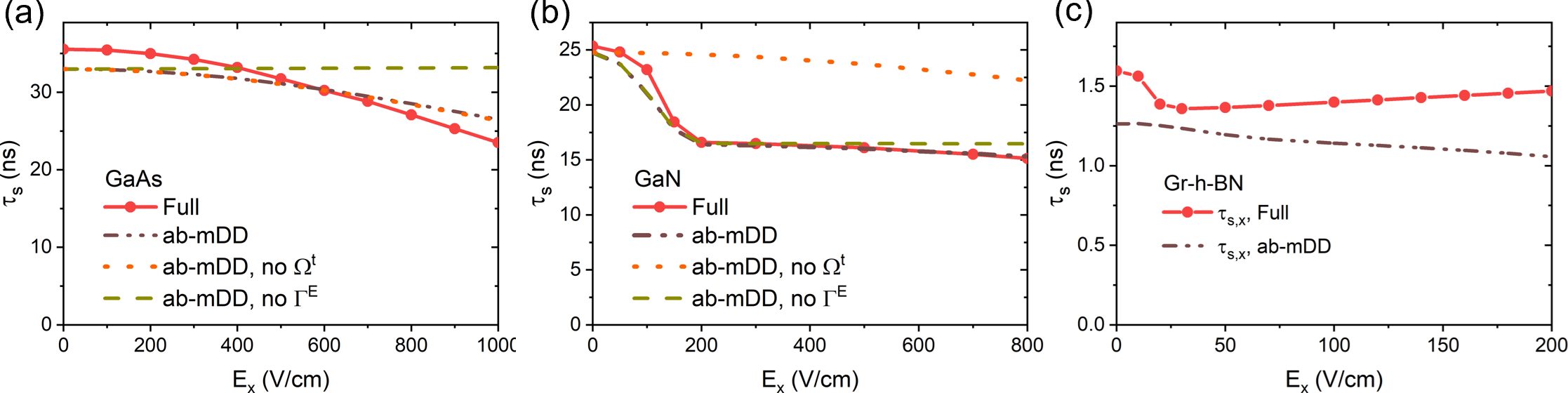}

\caption{Calculated $\tau_{s}(E_{x})$ of GaAs, GaN and graphene-$h$-BN by
solving full EVP and the ab-mDD model. ``no $\Gamma^{E}$'' or ``no
$\Omega^{t}$'' means that the $\Gamma^{E}$ (Eq. \ref{eq:GE}) or
$\Omega^{t}$ (Eq. \ref{eq:Omegat}) term of the model is not considered,
respectively.\label{fig:taus_abmDD}}
\end{figure*}

In contrast, $\tau_{s}$ of GaAs is strongly suppressed by an electric
field, decreasing by 34\% at 1000 V/cm {[}Fig. \ref{fig:taus}(b){]}.
Our theoretical results agree with earlier model calculations in Ref.
\citenum{jiang2009electron} (Fig. 15b therein). We find that $\tau_{sx}\left(E_{x}\right)$
of GaAs can be accurately described by 1-RR (1st-order RR, see Sec.
\ref{subsec:RR}) simulation and by a DP-like model derived from it.
This model gives (see Appendix K)
\begin{align}
\tau_{sx}^{-1}\left(E_{x}\right)= & \tau_{sx}^{-1}\left(E_{x}=0\right)+\tau\left\langle \Omega_{E,x}^{2}\right\rangle ,\label{eq:tausE_DP}
\end{align}
where $\tau$ is an effective lifetime comparable to carrier lifetime
(see Appendix K). Here, $\tau\left\langle \Omega_{E,x}^{2}\right\rangle $
is interpreted as electric-field-induced DP spin relaxation rate,
activated when spin precessions are present (see the end of Appendix
K). Numerically, $\left\langle \Omega_{E,x}^{2}\right\rangle $ can
be computed as $\left\langle U_{s\alpha}^{tL}\right|(L^{E})^{2}\left|U_{s\alpha}^{tR}\right\rangle $,
where $U_{s\alpha}^{tR(L)}$ is the right (left) eigenvector of $-L$($E_{x}$=0)
with corresponding eigenvalue being $\tau_{s,\alpha}^{-1}$($E_{x}$=0).

We then show $\tau_{s}\left(E_{x}\right)$ of GaN and graphene-$h$-BN
in Fig. \ref{fig:taus}(c)-(f). For GaN, $\tau_{s,x}$ and $\tau_{s,z}$
first converge to an intermediate value as $E_{x}$ rises to 200 V/cm,
then decrease slowly with $E_{x}$. Whereas $\tau_{s,y}$ decreases
monotonically with $E_{x}$, similar to the GaAs. According to our
prior work in Ref. \citenum{xu2025predicting}, the $E_{x}$-dependence
of $\tau_{s,\alpha}$ ($\alpha$=$x,y,z$) at $E_{x}$$\le$200 V/cm
is quite similar to the $B_{y}$-dependence ($B_{y}$ is external
magnetic field along $y$ direction) of $\tau_{s,\alpha}$ at $E_{x}$=0
but finite $B_{y}$. This is confirmed by our theoretical results
in Fig. \ref{fig:taus}(d). Therefore, our results indicate that the
electric field induces an effective global magnetic field along $y$
direction - ${\bf B}^{E_{x}}$, which is determined by spin-orbit
field ${\bf B}_{k}^{\mathrm{soc}}$ (see Appendix L),
\begin{align}
{\bf B}^{E_{x}}\approx & -\frac{e}{\hbar}E_{x}\tau\left\langle \frac{d{\bf B}_{k}^{\mathrm{soc}}}{dk_{x}}\right\rangle .\label{eq:BE}
\end{align}

In Rashba materials including GaN and graphene-$h$-BN, the main part
of ${\bf B}_{k}^{\mathrm{soc}}$ is $2\alpha\left(k_{y},-k_{x},0\right)$
with $\alpha$ Rashba coefficient, so that ${\bf B}^{E_{x}}$ is dominated
by $2e\hbar^{-1}E_{x}\tau\alpha$$\left(0,1,0\right)$. Indeed, using
our numerical values $\alpha$=0.018 eV/\AA and $\tau$=0.39 ps,
we find ${\bf B}^{E_{x}}$$\approx$0.11 Tesla at $E_{x}$=200 V/cm,
consistent with results in Fig. \ref{fig:taus}(d). Such ${\bf B}^{E_{x}}$
causes a global Larmor precession of spin observables along $x$ and
$z$ directions - $S_{x}$ and $S_{z}$, mixing their dynamics and
explaining the variations of $\tau_{s,x}$ and $\tau_{s,z}$ at $E_{x}$$\le$200
V/cm. The behavior is well captured by the 0\nobreakdash-RR simulation
{[}Fig \ref{fig:taus}(c){]}, in which the coupled dynamics of $S_{x}$
and $S_{z}$ in GaN yield (complex) spin relaxation rates
\begin{align}
\Gamma_{s\pm}\approx & \frac{\Gamma_{s,x}+\Gamma_{s,z}}{2}\pm\sqrt{\frac{\left(\Gamma_{s,x}-\Gamma_{s,z}\right)^{2}}{4}-\Omega_{xz}^{2}},\label{eq:G_0RRxz}
\end{align}
with
\begin{align}
\Gamma_{s,\alpha}= & \tau_{s,\alpha}^{-1}(E_{x}=0)-\left\langle U_{s\alpha}^{tL}\right|L^{E}\left|U_{s\alpha}^{tR}\right\rangle ,\label{eq:GiE_0RR}\\
\Omega_{xz}^{2}= & -\left\langle U_{sz}^{tL}\right|L^{E}\left|U_{sx}^{tR}\right\rangle \left\langle U_{sx}^{tL}\right|L^{E}\left|U_{sz}^{tR}\right\rangle .\label{eq:Omega2xz}
\end{align}

Since $L^{E}$ is in principle anti-Hermitian, $\Omega_{xz}^{2}$
is positive. For GaN, $\left\langle U_{s\alpha}^{tL}\right|L^{E}\left|U_{s\alpha}^{tR}\right\rangle $$\approx$0
and $\Gamma_{s,\alpha}$$\approx$$\tau_{s,\alpha}^{-1}(E_{x}=0)$
within the study $E$ range. Thus, when $E_{x}$ is large enough ($\ge$200
V/cm for GaN), $\Omega_{xz}^{2}\gg(\Gamma_{s,x}-\Gamma_{s,z})^{2}/4$,
we have
\begin{align}
\Gamma_{s\pm}\approx & \frac{\tau_{s,x}^{-1}(E_{x}=0)+\tau_{s,z}^{-1}(E_{x}=0)}{2}\pm i\sqrt{\Omega_{xz}^{2}},\label{eq:G_0RRxz_2}
\end{align}
leading to two same spin lifetimes $\tau_{s\pm}$=$1/(\mathrm{Re}\Gamma_{s\pm})$
for $x$ and $z$ directions.

At $E_{x}$$>$200 V/cm, the reduction of $\tau_{s,\alpha}$ with
$E_{x}$ is however not captured by 0-RR but well described by 1-RR.
This indicates such reduction arises from $E_{x}$-induced DP spin
relaxation $\tau\left\langle \Omega_{E,x}^{2}\right\rangle $.

In graphene-$h$-BN, the low\nobreakdash-field ($E_{x}$$\le$300
V/cm) behavior of $\tau_{s.\alpha}$ resembles that of GaN: $\tau_{s,x}$
and $\tau_{s,z}$ first decreases and increases, respectively, then
both vary slowly, whereas $\tau_{s,y}$ changes only gradually {[}Fig.
\ref{fig:taus}(e){]}. This again reflects the presence of a finite
${\bf B}^{E_{x}}$ along $y$ in this Rashba system.

\begin{figure*}
\includegraphics[scale=0.65]{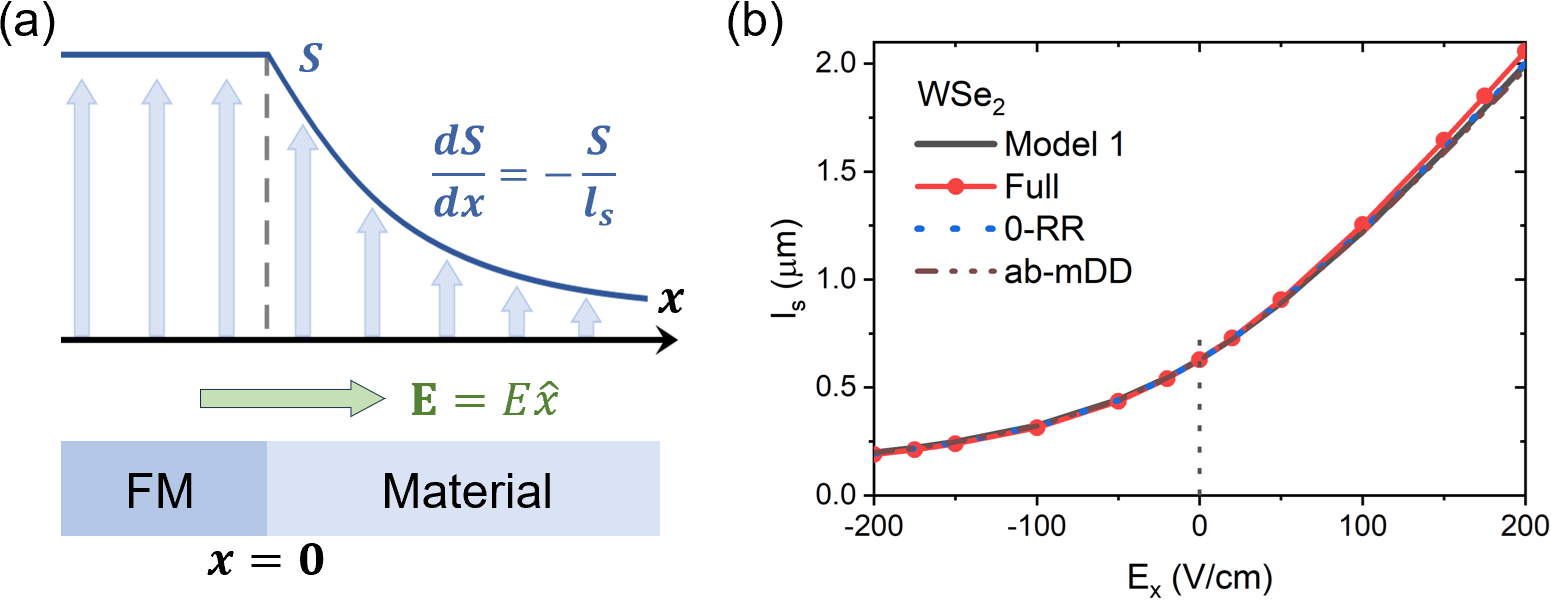}

\caption{(a) Schematic of the spin-diffusion setup. Spin are injected from
a ferromagnet (FM) at $x<0$ through a transparent interface into
the material at $x\ge0$, where they diffuse under zero or finite
${\bf E}$ along $x$ axis. (b) $E_{x}$ dependent spin diffusion
length ($l_{s}(E_{x})$) of holes of monolayer WSe$_{2}$ at 50 K
by different methods. ``Model 1'' uses the analytical formula Eq.
\ref{eq:model} with $\eta$ being $\eta_{1}$=$e/(v_{Fx}^{2}\widetilde{m}_{x})$
(Eq. \ref{eq:model1}). ``Full'' means that $l_{s}$ is calculated
by solving full EVP Eq. \ref{eq:gevp}. ``$n$-RR'' means that $l_{s}(E_{x})$
is computed by solving the reduced EVP Eq. \ref{eq:RR_EVP_ls} using
$n$th-order RR method with basis functions $V^{R(L)}$ given by Eq.
\ref{eq:RR_ls} (see Sec. \ref{subsec:RR}).\label{fig:wse2}}
\end{figure*}

Two notable differences appear in graphene\nobreakdash-$h$\nobreakdash-BN
compared with GaN. First, the electric-field-induced change of $\tau_{s,\alpha}$
arise not only from the explicit drift term in the master equation,
but also from the electric-field modification of $\rho^{\mathrm{eq}}$,
away from Fermi-Dirac function $f$ {[}Fig. \ref{fig:taus}(e){]}.
Second, $\tau_{s,x}$ and $\tau_{s,z}$ remain distinct over the whole
$E$ range and both slowly increase at higher $E_{x}$. As $\tau_{s,\alpha}(E_{x})$
of graphene-$h$-BN are accurately described by 0-RR simulations {[}Fig.
\ref{fig:taus}(f){]} and Eqs. \ref{eq:G_0RRxz}-\ref{eq:Omega2xz},
this fact can be explained as: in graphene-$h$-BN, $\left\langle U_{s\alpha}^{tL}\right|L^{E}\left|U_{s\alpha}^{tR}\right\rangle $
are complex values of small positive real parts and significant imaginary
part at higher $E_{x}$, which makes the $E_{x}$ effect very different
from a pure $B_{y}$\nobreakdash-field effect. The positive real parts
of $\left\langle U_{s\alpha}^{tL}\right|L^{E}\left|U_{s\alpha}^{tR}\right\rangle $
can slightly increase spin lifetimes, according to Eq. \ref{eq:GiE_0RR}.
Numerically, $(\Gamma_{s,x}-\Gamma_{s,z})^{2}/4$ is comparable to
$\Omega_{xz}^{2}$ within the studied $E$ range, so that the real
part of $\sqrt{(\Gamma_{s,x}-\Gamma_{s,z})^{2}/4-\Omega_{xz}^{2}}$
becomes non-negligible and $\mathrm{Re}\Gamma_{s+}\neq\mathrm{Re}\Gamma_{s-}$,
leading to two different spin lifetimes for $x$ and $z$ directions.

We further compare results of GaAs, GaN and graphene-$h$-BN by solving
full EVP and the ab-mDD model in Fig. \ref{fig:taus_abmDD}. The model
works well for GaN but yields a few percent errors for GaAs and significant
errors (up to 28\%) for graphene-$h$-BN. From theoretical results
of GaAs and graphene-$h$-BN in Figs. \ref{fig:taus} and \ref{fig:taus_abmDD},
the ab-mDD model appears worse than 1-RR simulation. This indicates
that 1-RR simulation with $V^{R(L)}$ set as Eq. \ref{eq:RR_taus}
probably approximates the $\rho$ solution more accurately than the
ab-mDD model, which is expected as 1-RR is exact at ${\bf E}$=0.

Moreover, as discussed in Sec. \ref{subsec:RR}, the $\Gamma^{E}$
(Eq. \ref{eq:GE}) and $\Omega^{t}$ (Eq. \ref{eq:Omegat}) terms
in the ab-mDD model represent the electric-field-induced DP spin relaxation
rate and temporal spin precession, respectively. Therefore, from the
comparison between the ab-mDD model and the version without the $\Gamma^{E}$
or $\Omega^{t}$ term in Fig. \ref{fig:taus_abmDD}(a)-(b), we conclude
that: (i) the $\tau_{s}$ reductions for GaAs and for GaN at $E_{x}$\textgreater 200
V/cm are both caused by the electric-field-induced DP spin relaxation;
(ii) the $\tau_{s}$ variation for GaN at $E_{x}$$\le$200 V/cm is
attributed to the electric-field-induced temporal spin precession;
and (iii) the $\Omega^{t}$ term has no effect on $\tau_{s}(E_{x})$
of GaAs, likely because the spin\nobreakdash-orbit fields in GaAs
are primarily of cubic\nobreakdash-Dresselhaus type, yielding ${\bf B}^{E_{x}}$$\approx$0
and hence no global field\nobreakdash-induced spin precession. These
conclusions are fully consistent with the analysis based on the 0\nobreakdash-RR
and 1\nobreakdash-RR simulations above.

\subsection{The electric-field effect on $l_{s}$}

\begin{figure*}
\includegraphics[scale=0.65]{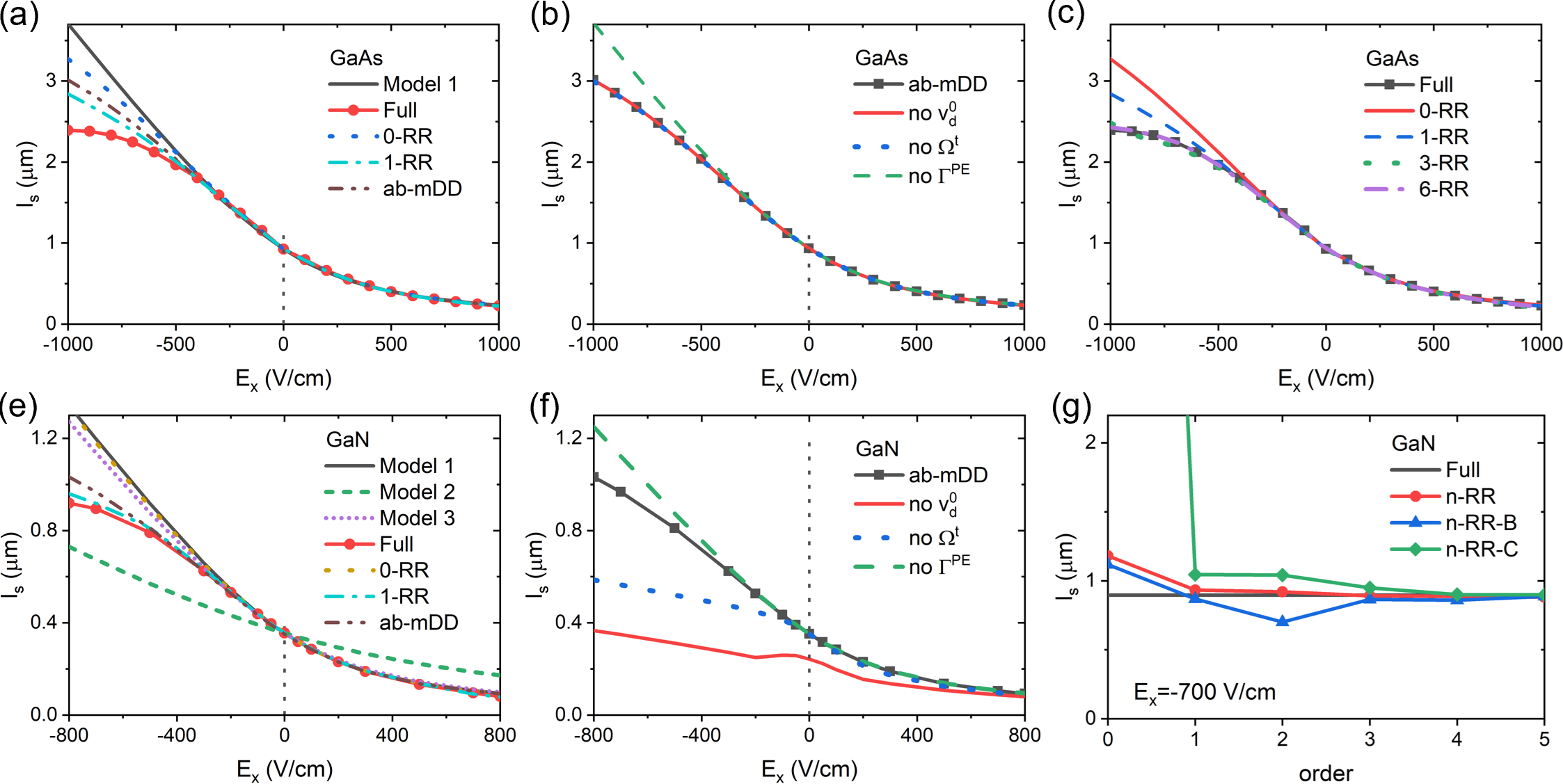}

\caption{$l_{s}(E_{x})$ of bulk GaAs at 300 K and bulk GaN at 100 K, computed
by different methods. Panel (g) shows $l_{s}$ of GaN at $E_{x}$=-700
V/cm computed by RR methods of different orders ($n$) and with different
types of basis functions $V^{R(L)}$, compared with results by solving
full EVP. ``Model 1, 2 and 3'' refer to the model $l_{s}(E_{x})$
formula Eq. \ref{eq:model} with the parameter $\eta$ set as $\eta_{1}$,
$\eta_{2}$=$\mu/D_{s}$ and $\eta_{3}$=$e/(k_{B}T)$ (Eqs. \ref{eq:model1}-\ref{eq:model3}),
respectively. ``no $v_{d}^{0}$'' corresponds to the ab-mDD model
without the $v_{d}^{0}$ term (Eq. \ref{eq:vd0}). $V^{R(L)}$ of
$n$-RR, $n$-RR-B and $n$-RR-C are given in Eqs. \ref{eq:RR_ls},
\ref{eq:RR-B} and \ref{eq:RR-C}, respectively.\label{fig:gaas_gan}}
\end{figure*}

We investigate the steady-state spin diffusion along $x$ direction,
as illustrated in Fig. \ref{fig:wse2}(a). Spins are injected from
a ferromagnet at $x<0$ to the semi-infinite material at $x\ge0$
and diffuses in it.

Fig. \ref{fig:wse2}(b) shows calculated $l_{s}(E_{x})$ of WSe$_{2}$
holes. For holes, $E_{x}>0$ and $E_{x}<0$ correspond to downstream
and upstream fields respectively. While for electrons, the situation
is reversed. Thus, it is clear that $l_{s}(E_{x})$ values of WSe$_{2}$
are significantly enhanced or suppressed by moderate downstream or
upstream fields respectively, as expected by the model formula Eq.
\ref{eq:model}. By comparing results by different methods, we find
that ``Model 1'' (Eq. \ref{eq:model} with $\eta$=$\eta_{1}$=$e/(v_{Fx}^{2}\widetilde{m}_{x})$
from Eq. \ref{eq:model1}), 0-RR simulation (Eqs. \ref{eq:RR_EVP_ls}-\ref{eq:RR_ls})
and the ab-mDD model all agree well with the full \textit{ab initio}
results {[}lines labeled as ``Full'' in Fig. \ref{fig:wse2}(b){]}.
``Model 2 and 3'', using Eq. \ref{eq:model} but with $\eta$ being
$\eta_{2}$=$\mu/D_{s}$ (Eq. \ref{eq:model2}) and $\eta_{3}$=$e/(k_{B}T)$
(Eq. \ref{eq:model3}), respectively, also yield accurate results
(not shown) for these materials. For EY systems such as WSe$_{2}$,
it has been expected that their spin diffusions are well described
by the standard drift-diffusion model.\citep{wu2010spin} Here, this
is confirmed by our full \textit{ab initio} and approximate simulations.

\begin{figure}
\includegraphics[scale=0.65]{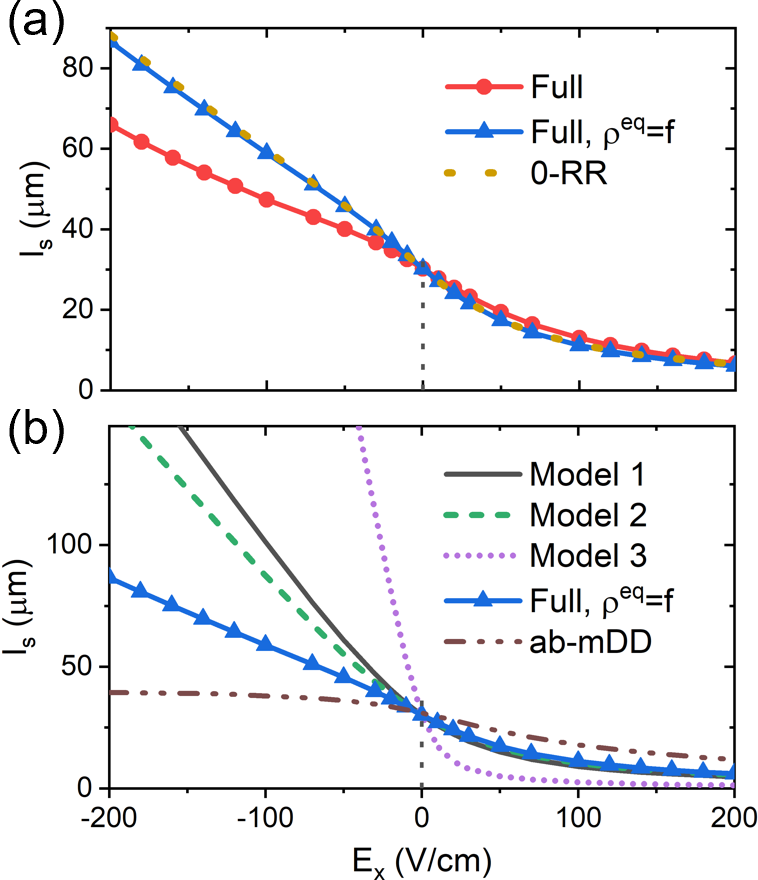}

\caption{$l_{s}(E_{x})$ of graphene-$h$-BN at 300 K, computed by different
methods.\label{fig:grbn}}
\end{figure}

Next, we examine $l_{s}(E_{x})$ of two typical DP systems - conduction
electrons of GaAs and GaN in Fig. \ref{fig:gaas_gan}. The full \textit{ab initio}
results {[}red solid lines in Fig. \ref{fig:gaas_gan}(a) and (d){]}
show that, as in WSe$_{2}$, a moderate downstream (upstream) field
strongly enhances (suppresses) $l_{s}$. However, the $l_{s}$ variation
with $E_{x}$ weakens at $E_{x}$$\le$-500 V/cm. At $E_{x}$$\ge$-400
V/cm, all approximate methods except ``Model 2'' reproduce the full
\textit{ab initio} results well. However, at $E_{x}$$\le$-500 V/cm,
``Model 1-3'' and 0-RR simulation all exhibit significant errors.
The ab-mDD model and 1-RR simulations substantially improve theoretical
results. The improvement is partly because of the inclusion of the
electric-field-induced DP spin relaxation, as discussed above. Overall,
1-RR simulation performs best, but it still has non-negligible errors
at $E_{x}$$\le$-800 V/cm for GaAs. Such errors can be reduced by
using higher-order RR simulations, as demonstrated in Fig. \ref{fig:gaas_gan}(c).
Our results indicate that for certain materials at high downstream
electric field, the $\rho$ solution for spin diffusion may not be
accurately expressed by relatively simple forms assumed in the ab-mDD
model (Eq. \ref{eq:rho_abmDD}), 0-RR and 1-RR simulations (Eq. \ref{eq:RR_ls});
instead, a more complete basis set provided by a higher\nobreakdash-order
RR simulation is required.

In Fig. \ref{fig:gaas_gan}(b) and (f), we show $l_{s}(E_{x})$ of
GaAs and GaN by the ab-mDD model and the version without the $v_{d}^{0}$,
$\Omega^{t}$ or $\Gamma^{E}$ term. For both GaAs and GaN, the $\Gamma^{E}$
term, corresponding to electric-field-induced DP spin relaxation,
reduces $l_{s}$ at high downstream electric field. The $\Omega^{t}$
term (corresponding to field-induced temporal spin precession) has
no effect on $l_{s}(E_{x})$ of GaAs but a significant effect on $l_{s}(E_{x})$
of GaN, similar to $\tau_{s}(E_{x})$ of them. The $v_{d}^{0}$ term,
corresponding to the zero-field spatial spin precession, strongly
affects $l_{s}(E_{x})$ of GaN in the whole $E_{x}$ range. The importance
of zero-field spatial spin precession on spin diffusion has been observed
in previous theoretical and experimental studies.\citep{xu2025predicting,hruvska2006effects,wu2010spin}
The effects of $v_{d}^{0}$, $\Omega^{t}$ and $\Gamma^{E}$ terms
are not considered in the simple model formula Eq. \ref{eq:model},
which explains the large errors of ``Model 1-3''.

\begin{table*}
\centering
\resizebox{0.99\textwidth}{!}{

\begin{tabular}{c|c|c|c|c|c|c|c}
\hline 
\multirow{2}{*}{} & \multirow{2}{*}{System} & \multirow{2}{*}{$\tau_{s}$ type} & \multirow{2}{*}{SOC fields} & \multicolumn{4}{c}{Relative error of method at $|E_{x}^{\mathrm{down}}|$=$E_{\mathrm{max}}$
(0.5$E_{\mathrm{max}}$; 0)}\tabularnewline
\cline{5-8} \cline{6-8} \cline{7-8} \cline{8-8} 
 &  &  &  & ``Model 1'' & ab-mDD & 0-RR & 1-RR\tabularnewline
\hline 
\hline 
\multirow{4}{*}{$\tau_{s}$} & WSe$_{2}$ & EY & -- & -- & 1\% (1\%; 1\%) & 1\% (0\%) & --\tabularnewline
\cline{2-8} \cline{3-8} \cline{4-8} \cline{5-8} \cline{6-8} \cline{7-8} \cline{8-8} 
 & GaAs & DP & Dresselhaus & -- & \textbf{13\%} (2\%; 7\%) & \textbf{52\%} (12\%) & 9\% (2\%)\tabularnewline
\cline{2-8} \cline{3-8} \cline{4-8} \cline{5-8} \cline{6-8} \cline{7-8} \cline{8-8} 
 & GaN & DP & Rashba & -- & 1\% (1\%; 2\%) & 10\% (2\%) & 1\% (0\%)\tabularnewline
\cline{2-8} \cline{3-8} \cline{4-8} \cline{5-8} \cline{6-8} \cline{7-8} \cline{8-8} 
 & Gr-BN & DP & Rashba & -- & \textbf{28\%} (\textbf{18\%}; \textbf{21\%}) & 1\% (0\%) & --\tabularnewline
\hline 
\multirow{4}{*}{$l_{s}$} & WSe$_{2}$ & EY & -- & 0\% (0\%) & 1\% (1\%; 0\%) & 0\% (0\%) & --\tabularnewline
\cline{2-8} \cline{3-8} \cline{4-8} \cline{5-8} \cline{6-8} \cline{7-8} \cline{8-8} 
 & GaAs & DP & Dresselhaus & \textbf{55\%} (9\%) & \textbf{26\%} (4\%; 1\%) & \textbf{37\%} (8\%) & \textbf{19\%} (2\%)\tabularnewline
\cline{2-8} \cline{3-8} \cline{4-8} \cline{5-8} \cline{6-8} \cline{7-8} \cline{8-8} 
 & GaN & DP & Rashba & \textbf{46\%} (10\%) & \textbf{12\%} (1\%; 1\%) & \textbf{45\%} (9\%) & 5\% (2\%)\tabularnewline
\cline{2-8} \cline{3-8} \cline{4-8} \cline{5-8} \cline{6-8} \cline{7-8} \cline{8-8} 
 & Gr-BN & DP & Rashba & \textbf{119\%} (\textbf{72\%}) & \textbf{55\%} (\textbf{35\%}; 3\%) & 2\% (1\%) & --\tabularnewline
\hline 
\end{tabular}}\caption{Relative errors of approximate methods compared with the full \textit{ab initio}
approach at selected downstream electric fields $E_{x}^{\mathrm{down}}$
for $\tau_{s}(E_{x})$ and $l_{s}(E_{x})$ of different systems, with
relevant system information. Gr-BN means graphene-$h$-BN. ``$\tau_{s}$
type'' indicates type or mechanism of spin relaxation. For each system,
$E_{\mathrm{max}}$ is the maximum studied $E_{x}$ of this system.
The value outside brackets is relative error at $|E_{x}^{\mathrm{down}}|$=$E_{\mathrm{max}}$.
The first number in brackets corresponds to $|E_{x}^{\mathrm{down}}|$=0.5$E_{\mathrm{max}}$.
The second number (if present) in brackets corresponds to $E_{x}$=0.
Errors exceeding 10\% are highlighted. By construction, $l_{s0}$
by ``Model 1'' and $n$-RR simulations are exactly the full \textit{ab initio}
$l_{s0}$ at $E_{x}$=0; therefore, their corresponding errors at
$E_{x}$=0 are omitted. Relative errors of ``Model 2 and 3'' are
not shown, since they are either similar to or larger than those of
``Model 1''.\label{tab:system_method_error}}
\end{table*}

We compare $l_{s}$ of GaN at $E_{x}$=-700 V/cm by $n$-RR, $n$-RR-B
and $n$-RR-C simulations with different $n$ in Fig. \ref{fig:gaas_gan}(f).
$n$-RR, $n$-RR-B and $n$-RR-C use different basis sets and set
$V^{R(L)}$ as Eqs. \ref{eq:RR_ls}, \ref{eq:RR-B} and \ref{eq:RR-C},
respectively (Sec. \ref{subsec:RR}). Among them, $n$-RR is the most
accurate, as expected because $n$-RR is exact at $E_{x}$=0. Notably,
0\nobreakdash-RR\nobreakdash-C fails completely, while 1\nobreakdash-RR\nobreakdash-C
yields much more reasonable result. This fact is also observed for
two other DP systems - GaAs and graphene-$h$-BN (not shown). On the
other hand, 0-RR-C gives reasonable results for a EY system - WSe$_{2}$.
For WSe$_{2}$ (at 50 K) at $E_{x}$=0 and 200 V/cm, $l_{s}$ by 0-RR-C
simulation is 0.58 and 1.7 $\mu$m respectively, not very far from
0.62 and 2 $\mu$m by solving full EVP. Therefore, while 0-RR-C may
work in some EY systems, it fails in DP systems. This probably indicates
that for spin diffusion in DP systems, the basis functions of 0-RR-C
lack essential components of the true solution of $\rho$, so that
it is necessary to introduce more basis functions via higher-order
RR-C. Overall, our results indicate that the choice of trial vectors
$V^{R(L)}$ is critical to RR simulations and higher-order RR simulations
generally improve accuracy.

We further show $l_{s}(E_{x})$ of graphene-$h$-BN at 300 K in Fig.
\ref{fig:grbn}. $E_{F}$ is set as 0.1 eV to ensure free carriers
are dominated by conduction electrons, so that spin-drift effect from
electrons is not reduced by that from holes. As for $\tau_{s}(E_{x})$
of graphene-$h$-BN, the electric-field-induced change of $l_{s}$
arise not only from the explicit drift term in the master equation,
but is also significantly affected by the electric-field modification
of $\rho^{\mathrm{eq}}$. This effect, not observed in WSe$_{2}$,
GaAs or GaN, further enhances the discrepancy between the full \textit{ab initio}
results and those by approximate methods in graphene-$h$-BN, since
this effect modifies the scattering term and cannot be directly considered
in approximate methods.

0-RR simulation is found in perfect agreement with the full \textit{ab initio}
simulation for $l_{s}(E_{x})$, just as it did for $\tau_{s}(E)$
of graphene-$h$-BN. This indicates that the electric-field-induced
DP spin relaxation (present in 1-RR but not in 0-RR) is absent in
graphene-$h$-BN within the studied $E_{x}$ range. As shown in Fig.
\ref{fig:grbn}(b), the ab-mDD model and ``Model 1-3'' all fail
to simulate $l_{s}(E_{x})$ correctly. ``Model 3'' performs worst,
since it relies on Einstein relation valid only in non-degenerate
semiconductors with parabolic bands. The failure of the ab-mDD model
is not surprising, since it already fail to simulate $\tau_{s}(E_{x})$
of graphene-$h$-BN. Our results suggest that certain important components
of the actual $\rho$ solution are correctly included in 0-RR simulation
but not in the ab-mDD model. While the ab\nobreakdash-mDD model may
be refined---for example, by decomposing $\rho$ into more than two
components---developing a simple yet universally accurate model is
non\nobreakdash-trivial and lies beyond the scope of this work.

Moreover, our numerical tests show that the failure of \textquotedblleft Model\,1--3\textquotedblright{}
and the ab\nobreakdash-mDD model persists in graphene--h\nobreakdash-BN
when an external magnetic field is applied, $E_{F}$ is shifted to
0.05\,eV, or the strength of spin-orbit field, scattering, or intervalley
scattering is rescaled (by a factor from 0.25 to 4). We have observed
that under particular sets of conditions, \textquotedblleft Model\,2\textquotedblright{}
yields $l_{s}(E_{x})$ values within 10\,\% of the full \textit{ab initio}
results. However, under the same conditions, ``Model 1 and 3'' and
the ab-mDD model still exhibit significant errors $\ge$20\%. Furthermore,
even this limited agreement with \textquotedblleft Model\,2\textquotedblright{}
disappears if the SOC or scattering strength is altered further, or
if external conditions are modified (for example, by applying a magnetic
field). These observations suggest that the apparent agreement of
\textquotedblleft Model\,2\textquotedblright{} is likely coincidental.
In conclusion, the drift-diffusion model generally does not work properly
in pristine graphene-$h$-BN. A more detailed microscopic analysis---potentially
combined with model\nobreakdash-Hamiltonian studies---will be needed
to fully understand the reasons for this systematic failure, a task
we leave for future work.

Overall, our theoretical results indicate that the simple model formulas
(``Model 1-3'') can cause significant errors of $l_{s}(E_{x})$
at moderate $E_{x}$. More sophisticated approximate methods, including
the ab-mDD model, 0-RR and 1-RR simulations, typically improve the
results but may still yield significant errors for certain materials
and/or conditions. Therefore, accurate simulations based on the full
\textit{ab initio} approach are critical to the correct prediction
and understanding of the electric-field effect on $l_{s}$ in general
cases.

Finally, to provide a clear overview of the accuracy of different
approximate methods, Table \ref{tab:system_method_error} summarizes
their relative errors in estimating $\tau_{s}(E_{x})$ and $l_{s}(E_{x})$
at selected downstream electric fields for each studied system, along
with relevant system information.

\section{Summary and Outlooks}

In summary, we implement the drift term due to finite ${\bf E}$ within
our \textit{ab initio} approach of $l_{s}$, based on a linearized
density-matrix master equation. Our method is applied to study the
electric-field effect on $\tau_{s}$ and $l_{s}$ in several representative
materials. We compare theoretical results by the full \textit{ab initio}
approach with several approximate methods, including low-order RR
method, our proposed ab-mDD model and a few simple model formulas
using Eq. \ref{eq:model} from the standard drift-diffusion model.
We find that although those approximate methods perform well in WSe$_{2}$,
they may cause significant errors for certain materials at moderate
$E_{x}$, which highlights the importance of carrying out full \textit{ab initio}
simulations. The detailed theoretical analysis reveal the importance
of a DP-type spin relaxation (Eq. \ref{eq:tausE_DP}) and a global
temporal spin precession due to $B^{E}$ (Eq. \ref{eq:BE}), both
induced by periodic-direction electric-field, to $\tau_{s}(E)$ and
$l_{s}(E)$. The effect of zero-magnetic-field spatial spin precession,
corresponding to the $v_{d}^{0}$ term in the ab-mDD model, is also
found critical to $l_{s}(E)$ of both GaN and graphene-$h$-BN. We
also find that the electric-field modification of $\rho^{\mathrm{eq}}$
can significantly affect $\tau_{s}(E)$ and $l_{s}(E)$, an effect
absent in previous theoretical studies.

Our \textit{ab initio} framework with finite ${\bf E}$ may be further
extended to simulate other linear or nonlinear carrier and spin transport
properties, with quantum treatment of electron scattering processes
and the inherent inclusion of Berry phase effects. In the present
work, the scattering term does not explicitly include ${\bf E}$,
which is however an approximation. A more complete theory would require
reformulating the density-matrix master equation to fully account
for the field dependence of scattering, for instance, by adopting
a phonon-assisted density matrix formalism\citep{iotti2017phonon}.
This extension, however, would introduce considerable complexity,
notably through the appearance of covariant derivatives of the e-ph
matrix elements. Given the considerable technical challenges involved
and the expectation that this contribution is weaker than the electric-field
effect due to the drift term, we defer the treatment of field-dependent
scattering to future work.

\section*{Acknowledgments}

This work is supported by National Natural Science Foundation of China
(Grant No. 12574257 and 12304214), Fundamental Research Funds for
Central Universities (Grant No. JZ2023HGPA0291 and JZ2025HGQA0310).
This research used resources of the HPC Platform of Hefei University
of Technology.

\section*{Appendices}

\subsection*{Appendix A: $L^{v_{j}}$ and $L^{e}$ in linearized master equation}

$L^{v_{j}}$ and $L^{e}$ are
\begin{align}
L_{kab,k'cd}^{v_{j}}= & \frac{1}{2}\left(v_{j,kac}\delta_{bd}+\delta_{ac}v_{j,kdb}\right)\delta_{kk'},\\
L_{kab,k'cd}^{e}= & \frac{-i}{\hbar}\left(H_{kac}^{e}\delta_{bd}-\delta_{ac}H_{kdb}^{e}\right)\delta_{kk'}.
\end{align}

\subsection*{Appendix B: Comparisons with standard Boltzmann transport equation
(BTE) and kinetic spin Bloch equation (KSBE)}

Quantum master equation can be viewed as a generalization of standard
BTE and is reduced to BTE by replacing $\rho_{kab}$ to its diagonal
part $\rho_{ka}^{d}$=$\rho_{kab}\delta_{ab}$. Define $P_{ka,k'b}^{d}$=$P_{kab,k'bb}$,
the resulting scattering term $C[\rho^{\mathrm{tot},d}]$ reads
\begin{align}
C_{ka}[\rho^{\mathrm{tot},d}]= & (1-\rho_{ka}^{\mathrm{tot},d})P_{ka,k'b}^{d}\rho_{k'b}^{\mathrm{tot},d}\nonumber \\
 & -\rho_{ka}^{\mathrm{tot},d}P_{ka,k'b}^{d}(1-\rho_{k'b}^{\mathrm{tot},d}),
\end{align}
which is derived using the property of $P_{ka,k'b}^{d}$=$P_{k'b,ka}^{d}$
and all $P^{d}$ elements are real values. $P_{ka,k'b}^{d}$ is actually
the e-ph transition rate between electron states $\{k,a\}$ and $\{k',b\}$.

The linearized scattering term $L^{Cd}$ reads
\begin{align}
L_{ka,k'b}^{Cd}= & -P_{ka,k'b}^{d}(\rho_{ka}^{d}-\rho_{k'b}^{d}),
\end{align}
which has been commonly employed in theoretical simulations based
on standard BTE.

A critical difference between quantum master equation and standard
BTE is that Larmor spin precession is not directly included in standard
BTE.

Standard KSBE is an alternative form of quantum master equation for
two-band systems. Within standard KSBE, the 2$\times$2 $\rho$ is
expressed as
\begin{align}
\rho_{k}= & \frac{\widetilde{n}_{k}}{2}I+\widetilde{s}_{k\alpha}\sigma_{\alpha},
\end{align}
where $\widetilde{n}_{k}$=$\text{\ensuremath{\rho_{k11}}+\ensuremath{\rho_{k22}}}$,
$\widetilde{s}_{kx}$=$\mathrm{Re}\rho_{k12}$, $\widetilde{s}_{ky}$=$-\mathrm{Im}\rho_{k12}$
and $\widetilde{s}_{kz}$=$(\rho_{k11}-\rho_{k22})/2$.

In principle, standard KSBE and quantum master equation are equivalent
for two-band systems. Nevertheless, our used quantum master equation
is more general, as it applies to arbitrary band structures. Moreover,
practical forms of KSBE in the literature may employ additional approximations.
For instance, the energy-conservation-related function in the scattering
term may use electron energies without considering the modifications
due to spin-orbit fields. This is usually a minor assumption but may
cause some errors if spin-orbit fields are very strong. Additionally,
previous KSBE simulations relied on model Hamiltonian and highly-simplified
treatment of the scattering term, these limitations are removed in
our \textit{ab initio} quantum-master-equation simulations.

\subsection*{Appendix C: Spin perturbative density-matrix $\rho^{s_{\alpha}}$}

$\rho^{s_{\alpha}}$ reads
\begin{align}
\rho_{kab}^{s_{\alpha}}= & c^{\rho}\left(\frac{\Delta f}{\Delta\epsilon}\right)_{kab}s_{\alpha,kab},\label{eq:rhos}\\
\left(\frac{\Delta f}{\Delta\epsilon}\right)_{kab}= & \left(\frac{df}{d\epsilon}\right)_{kab}\delta_{\epsilon_{ka}\epsilon_{kb}}+\frac{f_{ka}-f_{kb}}{\epsilon_{ka}-\epsilon_{kb}}\left(1-\delta_{\epsilon_{ka}\epsilon_{kb}}\right),
\end{align}
where $c^{\rho}$ is an arbitrary constant.

\subsection*{Appendix D: The computation of the observable evolution and the evaluation
of $\tau_{s}$ and $l_{s}$}

Having the observable operator $o_{\kappa}$, and defining $o_{\mu}^{t}$=$N_{k}^{-1}$$\sum_{\kappa}o_{\kappa}^{*}U_{\kappa\mu}^{tR}$
and $o_{\nu}^{x}$=$N_{k}^{-1}$$\sum_{\kappa}o_{\kappa}^{*}U_{\kappa\nu}^{xR}$,
the observable evolution for relaxation problem is:\citep{xu2025predicting}
\begin{align}
O(t)= & \mathrm{Re}(\Sigma_{\mu}c_{\mu}^{t}o_{\mu}^{t}e^{-\Gamma_{\mu}t}),\label{eq:Ot}
\end{align}
and for diffusion problem is:
\begin{align}
O(x)= & \mathrm{Re}(\Sigma_{\nu}c_{\nu}^{x}o_{\nu}^{x}e^{-\lambda_{\nu}^{x}x}).\label{eq:OX}
\end{align}

Eq. \ref{eq:Ot} and \ref{eq:OX} accurately describes the observable
decay for given $\rho^{\mathrm{pert}}$. The observable dynamics consist
of dynamics of individual decay modes, highly simplifying the analysis
compared to previous real-time method\citep{xu2024spin,xu2024graphite}.
For given $\rho^{\mathrm{pert}}$ and $o$, the relevance of a mode
to $O(t)$ or $O(x)$ is determined by its eigenvectors and can be
measured by the corresponding mode-resolved normalized relevance factor
\begin{align}
\mathscr{R}_{\mu}^{t(x)}= & \frac{\left|c_{\mu}^{t(x)}o_{\mu}^{t(x)}\right|}{\mathrm{max}_{\mu'}\left\{ \left|c_{\mu'}^{t(x)}o_{\mu'}^{t(x)}\right|\right\} }.
\end{align}

For slow decay of spin observable, typically only a few modes are
relevant, so that $O(t)$ or $O(x)$ is well described by eigenvalues
and eigenvectors of these ``relevant'' modes. If only one mode $\mu_{0}$
is relevant, diffusion length along $x$ direction - $l^{x}$ (lifetime
$\tau$) for the observable is simply $l^{x}$ ($\tau$) of this $\mu_{0}$
mode. With multiple non-degenerate relevant modes, to describe the
observable decay by a single $l^{x}$ ($\tau$) value, we need to
define an effective or averaged diffusion length $\overline{l^{x}}$
(lifetime $\overline{\tau}$) length as
\begin{align}
\overline{\tau}= & \frac{\sum_{\mu\in\mu^{\mathrm{rel}}}\mathscr{R}_{\mu}^{t}\tau_{\mu}}{\sum_{\mu\in\mu^{\mathrm{rel}}}\mathscr{R}_{\mu}^{t}},\\
\overline{l^{x}}= & \frac{\sum_{\nu\in\nu^{\mathrm{rel}}}\mathscr{R}_{\nu}^{x}l_{\nu}^{x}}{\sum_{\nu\in\nu^{\mathrm{rel}}}\mathscr{R}_{\nu}^{x}},
\end{align}
where $\mu^{\mathrm{rel}}$ ($\nu^{\mathrm{rel}}$) includes the ``relevant''
modes with $\mathscr{R}_{\mu}^{t}\ge$0.1 ($\mathscr{R}_{\nu}^{x}\ge$0.1).
In this work, for $s_{x}$ spin perturbation and $s_{x}$ observable
operator, $\mu^{\mathrm{rel}}$ ($\nu^{\mathrm{rel}}$) includes either
one or two modes.

\subsection*{Appendix E: About Ansatz III for the ab-mDD model}

The reasons why Ansatz III for Eq. \ref{eq:proj_ldmme} typically
holds are:

First, for $L^{P}\rho^{P}$=$\left(L^{C}+L^{e,P}+L^{E,P}\right)\rho^{P}$,
we have $|L^{C}\rho^{P}|$$\approx$$\Gamma_{\alpha}^{\mathrm{EY}}$$|\rho^{P}|$.
Both $L^{e,P}$ and $L^{E,P}$ are approximately anti-Hermitian matrices,
so that $|\left(L^{e,P}+L^{E,P}\right)\rho^{P}|$$\approx$$|\Omega|$$|\rho^{P}|$
with $\Omega$ being an effective Larmor precession frequency. As
$\Gamma_{\alpha}^{s}$ and $|\Omega|$ are typically tiny, $|L^{P}\rho^{P}|$
is tiny and much smaller than $|L^{C,Q}\delta|$ and the left 3rd
term. $d\rho^{P}/dt$ should have similar magnitude to $L^{P}\rho^{P}$
and therefore likewise tiny.

Second, $L^{C,Q}\delta$$\approx$$\tau_{m}^{-1}$$|\rho^{P}|$ with
$\tau_{m}$ being momentum lifetime. Typically, $\tau_{m}^{-1}$ is
relatively larger, considering that $\delta\ll\rho_{\kappa}^{P}$,
we can safely assume that the left 4th term is much smaller than $|L^{C,Q}\delta|$
and the left 3rd term.

\subsection*{Appendix F: The derivation of Eq. \ref{eq:ab-mDD}}

From Ansatz III for the ab-mDD model, we obtain the following two
equations:
\begin{align}
\left[L^{{\bf v}}\cdot\nabla_{{\bf R}}-\left(L^{e,Q}+L^{E,Q}\right)\right]\rho^{P}\approx & L^{C,Q}\delta,\label{eq:condition1}\\
\frac{d\rho^{P}}{dt}+\left[L^{{\bf v}}\cdot\nabla_{{\bf R}}-\left(L^{e}+L^{E}\right)\right]\delta\approx & L^{P}\rho^{P}.\label{eq:condition2}
\end{align}

From Eq. \ref{eq:condition1},
\begin{align}
\delta\approx & \left(L^{C,Q}\right)^{-1}\left[L^{{\bf v}}\cdot\nabla_{{\bf R}}-\left(L^{e,Q}+L^{E,Q}\right)\right]\rho^{P}.
\end{align}

Insert it into Eq. \ref{eq:condition2} and multiply $\varrho^{s,\dagger}$
to the left, we obtain Eq. \ref{eq:ab-mDD}.

\subsection*{Appendix G: The approximate formula of effective mobility $\widetilde{\mu}_{j}$
in Eq. \ref{eq:muj_tilde} of the ab-mDD model\label{subsec:model_from_RR}}

Suppose spins are highly polarized along $\alpha$ direction or there
are two Kramers degenerate bands with one spin-up and the other spin-down.
In these cases, spin matrix (in Bloch basis) $s_{\alpha}$ is approximately
$k$-independent. Therefore, we have
\begin{align}
\rho_{k}^{s_{\alpha}}= & c^{\rho}\frac{df_{k}}{d\epsilon}s_{\alpha},
\end{align}

We also suppose $\varrho^{s_{\alpha}}$=$c^{\varrho}s_{\alpha}$.
$c^{\rho}$ and $c^{\varrho}$ are real constants and their choices
must satisfy $\varrho^{s_{\alpha}}\rho^{s_{\alpha}}$=1.

Approximate $T$ as momentum lifetime $\tau_{m}$, we have
\begin{align}
\widetilde{\mu}_{j}\approx & -\tau_{m}E_{j}^{-1}\varrho^{s_{\alpha}}\left(L^{v_{j}}L^{E,Q}+L^{E}L^{v_{j}}\right)\rho^{s_{\alpha}}.
\end{align}

Suppose $\xi$-part of $L^{E}$ is negligible, which is satisfied
in one-band systems and in many systems if smooth basis functions
are chosen. Thus, we can approximate $L^{E}\approx\frac{eE_{j}}{\hbar}\frac{d}{dk_{j}}$.
Therefore, considering that $L^{E,Q}\approx L^{E}$, we have
\begin{align}
E_{j}^{-1}\varrho^{s_{\alpha}}L^{v_{j}}L^{E}\rho^{s_{\alpha}}= & \frac{e}{\hbar}c^{\varrho}c^{\rho}\mathrm{Tr}\left(s_{\alpha}^{2}\right)\int_{\Omega}d{\bf k}v_{kj}\frac{d}{dk_{j}}\frac{df_{k}}{d\epsilon}\nonumber \\
= & -\frac{e}{\hbar}c^{\varrho}c^{\rho}\int_{\Omega}d{\bf k}\frac{df_{k}}{d\epsilon}\frac{d}{dk_{j}}v_{kj}\nonumber \\
= & -e\varrho_{s}^{\dagger}\frac{dv_{kj}}{\hbar dk_{j}}\rho_{s},\\
E_{j}^{-1}\varrho^{s_{\alpha}}L^{E}L^{v_{j}}\rho^{s_{\alpha}}= & 2\frac{e}{\hbar}c^{\varrho}c^{\rho}\int_{\Omega}d{\bf k}\frac{d}{dk_{j}}\left(v_{kj}\frac{df_{k}}{d\epsilon}\right)\nonumber \\
= & 0.
\end{align}

Define
\begin{align}
\widetilde{m}^{-1}= & \varrho_{s}^{\dagger}\frac{dv_{kx}}{\hbar dk_{x}}\rho_{s},\label{eq:mass}
\end{align}
we then have
\begin{align}
\widetilde{\mu}\approx & e\tau_{m}/\widetilde{m}.
\end{align}

\subsection*{Appendix H: The solution for 1D steady-state diffusion within the
ab-mDD model}

For 1D steady-state diffusion along $x$ direction and assuming ${\bf E}||x$,
we have
\begin{align}
D_{xx}\frac{d^{2}S}{dx^{2}}+v_{x}^{d}\frac{dS}{dx}= & \Gamma^{P}S.
\end{align}

The general solution is
\begin{align}
S\left(x\right)= & Y^{R}e^{-\lambda x}c^{x},
\end{align}
with $\lambda$ and $Y^{R}$ satisfy
\begin{align}
D_{xx}Y^{R}\lambda^{2}-v_{x}^{d}Y^{R}\lambda= & \Gamma^{P}Y^{R}.\label{eq:QEP}
\end{align}

This is quadratic eigenproblem. Define $Z^{R}=\left(Y^{R},\lambda Y^{R}\right)$,
we have the standard eigenproblem for $\lambda$:
\begin{align}
A^{Z}Z^{R}= & Z^{R}\lambda,\\
A^{Z}= & \left(\begin{array}{cc}
I & 0\\
0 & D_{xx}^{-1}
\end{array}\right)\left(\begin{array}{cc}
0 & I\\
\Gamma^{P} & v_{x}^{d}
\end{array}\right)\nonumber \\
= & \left(\begin{array}{cc}
0 & I\\
D_{xx}^{-1}\Gamma^{P}, & D_{xx}^{-1}v_{x}^{d}
\end{array}\right).
\end{align}

For 1D diffusion at $x\ge0$ with the boundary conditions:
\begin{align}
S\left(0\right)= & S_{0},\\
S\left(+\infty\right)= & 0,
\end{align}
where $S_{0}$ can be determined by maximum the spin observable. Then,
we have
\begin{align*}
c^{x}= & Y^{R,-1}S_{0},
\end{align*}
and
\begin{align}
S_{m}\left(x\right)= & \sum_{\lambda_{\mu}>0,\kappa'}Y_{m\mu}^{R}e^{-\lambda_{\mu}x}\left(Y^{R,-1}\right)_{\mu n}S_{0,n}.
\end{align}

If $V^{sR}$ has only one column and is highly relevant to spin, then
$A^{Z}$ is 2$\times$2 matrix. Assuming $\Gamma^{P}$=$\tau_{s0}^{-1}$,
the eigenvalues are
\begin{align}
\lambda_{\pm}= & \frac{D_{xx}^{-1}v_{x}^{d}}{2}\pm\sqrt{\left(\frac{D_{xx}^{-1}v_{x}^{d}}{2}\right)^{2}+D_{xx}^{-1}\Gamma^{P}}\nonumber \\
= & \frac{E}{2}\frac{\widetilde{\mu}_{x}}{D_{xx}}\pm\sqrt{\frac{E^{2}}{4}\left(\frac{\widetilde{\mu}_{x}}{D_{xx}}\right)^{2}+\frac{1}{l_{s0}^{2}}},\\
l_{s0}= & \sqrt{D_{xx}\tau_{s0}},
\end{align}
which is the same as Eq. \ref{eq:model} from the standard drift-diffusion
model.

\subsection*{Appendix I: Computational details}

Our computational setups are the same as in our prior paper\citep{xu2025predicting}
(some data of this paper are openly available). The ground-state electronic
structure, phonons, as well as the e-ph matrix elements are firstly
calculated using DFT with relatively coarse $k$ and $q$ meshes in
the open-source DFT plane-wave code JDFTx\citep{sundararaman2017jdftx}.
For GaAs, WSe$_{2}$ and graphene-$h$-BN, the lattice constants are
taken 5.653, 3.32 and 2.465 $\text{Å}$ respectively, as in our previous
work\citep{xu2021ab,habib2022electric}. For GaN, we use relaxed lattice
constants with $a$=3.103 $\text{Å}$ and $c$=5.044 $\text{Å}$.
We use PBE exchange-correlation functional\citep{perdew1996generalized},
except that SCAN functional\citep{sun2015strongly} is used for GaAs
as in our previous work\citep{xu2021ab}. We use Optimized Norm-Conserving
Vanderbilt (ONCV) pseudopotentials\citep{hamann2013optimized} with
spin-orbit coupling. The plane-wave cutoff energies are 34, 64, 74
and 74 Ry for GaAs, WSe$_{2}$, graphene-$h$-BN and GaN respectively.
For graphene-$h$-BN, the DFT+D2 correction method\citep{grimme2006vdw}
with scale factor $s_{6}=0.5$ is used, to be consistent with our
previous work.\citep{habib2022electric}. The DFT calculations use
$24^{3}$, $12^{2}$, $28\times28\times20$ and $24^{2}$ $k$ meshes
for GaAs, WSe$_{2}$, GaN and graphene-$h$-BN respectively. The phonon
calculations employ $4^{3}$, $6^{2}$, $4^{3}$ and $6^{2}$ supercells
respectively through finite difference calculations.

We then transform all quantities from plane wave basis to maximally
localized Wannier function basis\citep{marzari1997maximally} and
interpolate them\citep{giustino2017electron} to substantially finer
k and q meshes. For WSe$_{2}$, GaAs, GaN and graphene-$h$-BN, the
Wannierizations use 22, 16, 24 and 54 Wannier functions, respectively.
The inner (or frozen) energy windows include 14, 8, 12 and 16 valence
bands, respectively. The upper bounds of inner windows are 2.5 eV,
2.9 eV, 6.6 eV and 14 eV higher than CBM (conduction band minimum),
respectively. The outer (disentanglement) energy windows are tens
of eV wider than the inner ones. For each material, we do Wannier
fittings starting from tens of or hundreds of initial sets of random
Wannier centers, and select one set having smallest errors in key
quantities for spin decay---band\nobreakdash-edge energy splittings,
spin\nobreakdash-mixing parameters, and spin--orbit fields. The Wannier
interpolation approach fully accounts for polar terms in the e-ph
matrix elements and phonon dispersion relations, using the approach
of Refs. \citenum{verdi2015frohlich} and \citenum{sohier2017breakdown}
for the 3D and 2D systems.

For the relaxation and diffusion problems, we only include electronic
states close to VBM energy $E^{\mathrm{vbm}}$, CBM energy $E^{\mathrm{cbm}}$
or Fermi level $E_{F}$. More explicitly, we include states within
the energy window $[E^{\mathrm{ref}}-n^{w}k_{B}T,E^{\mathrm{ref}}+n^{w}k_{B}T]$,
where $E^{\mathrm{ref}}$ is $E^{\mathrm{vbm}}$ or $E^{\mathrm{cbm}}$
for semiconductors and is $E_{F}$ for metallic systems. $n^{w}$
is typically 7-14, depending on material, temperature, etc. Typically,
there are about 2000-7000 states being included for the calculations
of $\tau_{s}$ and $l_{s}$. For example, for graphene-$h$-BN, the
fine $k$ and $q$ meshes are $480\times480$ and $n^{w}$=10, leading
to $\approx$2200 states within the energy window. We have checked
the convergence of $\tau_{s}$ and $l_{s}$ with the number of $k$-points
and $n^{w}$, and the convergence errors are found within 10$\%$.
The energy-conservation smearing parameter $\sigma_{G}$ is chosen
to be comparable or smaller than $k_{B}T$ for each calculation, e.g.,
$\sigma_{G}$ is typically 0.02 eV for 300 K. $\tau_{s}$ and $l_{s}$
are computed by our python script using physical quantities at fine
meshes.

Overall computational demands are moderate. Most jobs run on one or
two Intel Xeon Gold 6248R (48 cores) or 6348 (56 cores) processors.
Phonon calculations require a few to tens of hours. Wannier fitting,
depending on the number of initial trial sets, totally takes from
a few minutes to tens of hours. A single Wannier interpolation typically
completes in tens of minutes to a few hours. Each spin\nobreakdash-decay
simulation (at a given electric field) by our python script requires
between a few minutes and three hours, strongly dependent on the chosen
k-point density and number of bands.

\subsection*{Appendix J: About comparison with experiments}

\begin{figure}
\includegraphics[scale=0.3]{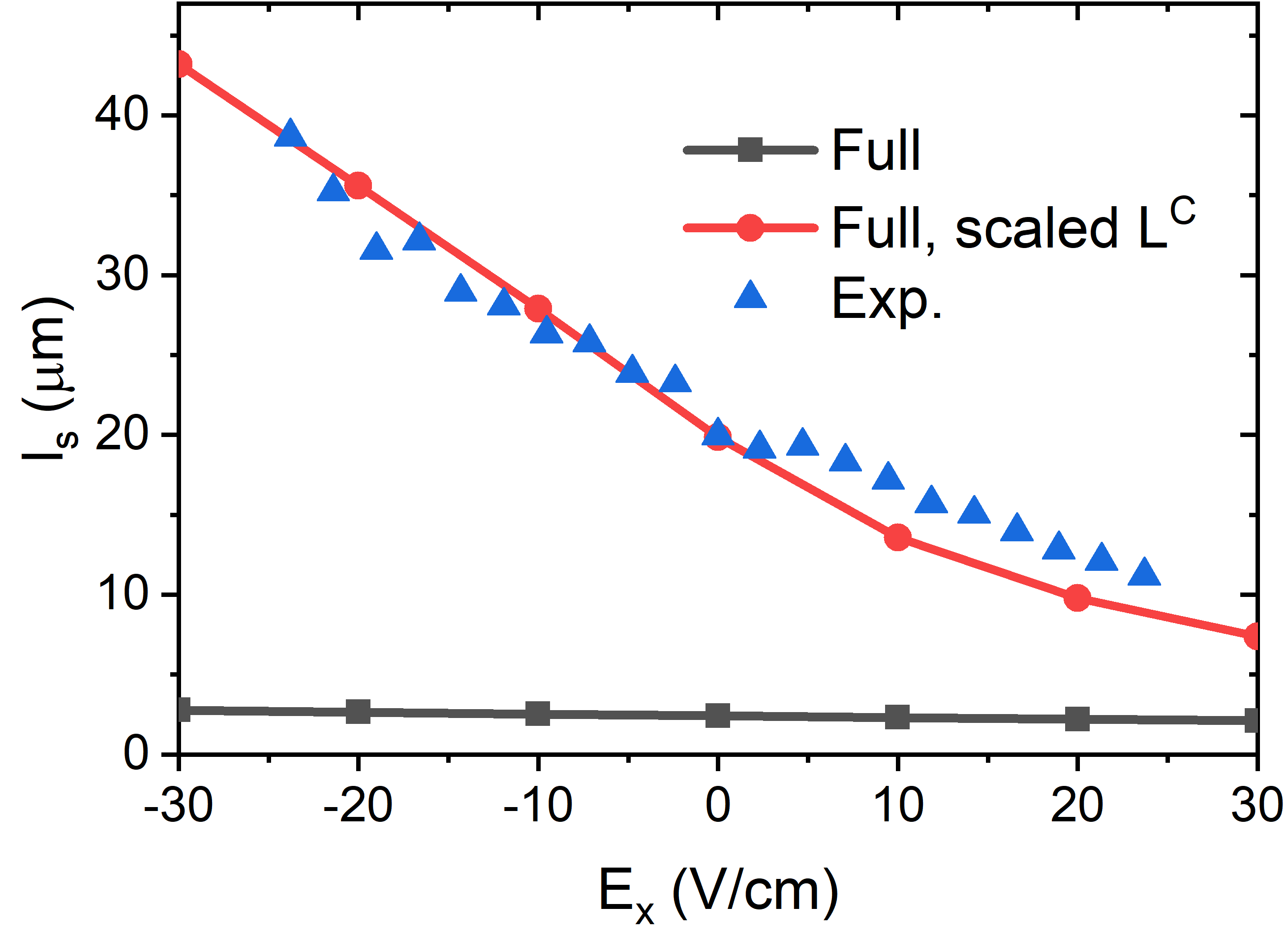}

\caption{Calculated $l_{s}(E_{x})$ of electrons of bulk germanium (Ge) at
300 K by solving full EVP, compared with experimental data from Ref.
\citenum{zucchetti2022electric} measured using inverse spin-Hall
effect. ``scaled $L^{C}$'' means that the scattering term is scaled
by a constant to fit the experimental value of $l_{s0}$. Note that
experimental ranges of room-temperature $\tau_{s0}$ and $l_{s0}$
of Ge across different papers are both quite large\citep{hamaya2018spin,rinaldi2016determination,zucchetti2019doping,yamada2021experimental}
- experimental $\tau_{s0}$ ranges from 0.25 ns to 132 ns, while experimental
$l_{s0}$ ranges from less than 0.5 $\mu$m to 34 $\mu$m. Our predicted
$\tau_{s0}$ and $l_{s0}$ are 0.64 ns and 2.4 $\mu$m (without a
scaling factor of $L^{C}$), within the experimental ranges but much
shorter than values from Refs. \citenum{zucchetti2022electric} (giving
$l_{s0}$ of 20 $\mu$m) and \citenum{zucchetti2019doping} (giving
$\tau_{s0}$ of 132 ns and $l_{s0}$ of 34 $\mu$m).\label{fig:ge}}
\end{figure}

In this work, we carry out theoretical studies of the electric-field
effects on $\tau_{s}$ and $l_{s}$ of a few relatively simple materials.
The comparisons with experiments are not simple due to several issues:

(i) We consider spin decay only due to only e-ph scattering. However,
in realistic samples, impurity scattering can be crucial. In fully
\textit{ab initio} simulations, the inclusion of impurity is highly
demanding, requiring large supercell calculations and precise knowledge
of impurity types and concentrations.

(ii) The experimental materials may be too difficult to be modeled
in full \textit{ab initio} simulations. For instance, fully \textit{ab initio}
modelings of quantum well, quantum wire and graphene on SiO$_{2}$,
studied in some experimental studies\citep{kwon2008electric,pramanik2003spin,jozsa2008electronic},
require extremely large supercells. Moreover, our studied systems
may be quite different from those in experiments, in carrier density,
substrate, temperature, etc.

(iii) Both theoretical and experimental data may have significant
uncertainties or errors. Theoretical results are limited by the accuracy
of DFT with specific functionals, while experimental values are sensitive
to various factors such as sample quality, measurement technique and
data analysis method. For instance, the experimental values of room-temperature
$l_{s0}$ of graphene samples, GaN and Pt, RuO$_{2}$ range from a
few to 31 $\mu$m, 135-510 nm, 1-11 nm and 2-12 nm, respectively,
in the literature\citep{avsar2020colloquium,wesselink2019calculating,xu2025predicting}.

Given that the primary goals of this work are to establish a fully
\textit{ab} framework for computing $\tau_{s}(E)$ and $l_{s}(E)$
and to demonstrate its application to a few representative systems
while validating various approximate methods, we would like to avoid
the complexities above, which may introduce confusions of the readers.
Therefore, we do not present direct comparisons with experimental
data in the main text.

Overall, there exist limited number of experimental studies of the
effect of periodic-direction electric field on $\tau_{s}$ or $l_{s}$.
For $l_{s}$, to the best of our knowledge, there exist only three
papers (Refs. \citenum{kwon2008electric}, \citenum{jozsa2008electronic}
and \citenum{zucchetti2022electric}) directly presenting experimental
data of $l_{s}$ as a function of periodic-direction electric field.
Since the materials in Refs. \citenum{kwon2008electric} and \citenum{jozsa2008electronic}
are difficult to be simulated, we will try to study a relatively simple
material in Ref. \citenum{zucchetti2022electric} - bulk germanium
(Ge).

Standard PBE functional incorrectly predicts Ge to be metallic, whereas
it is know that Ge is a semiconductor of band gap about 0.66 eV and
CBM located at $L$ point. Therefore, we select the MGGAC functional\citep{MGGAC,MGGAC_bp},
as it correctly captures the semiconducting nature of Ge, yielding
a band gap of 0.94\,eV and the CBM at $L$ point. Furthermore, MGGAC
provides lattice constants in agreement with both experimental and
PBE values. This ensures the accuracy of force simulation with MGGAC,
which is crucial for phonon and electron-phonon simulations. Hybrid
functionals are not used due to their prohibitive computational cost,
which makes subsequent phonon and electron-phonon calculations infeasible.

With MGGAC functional, we simulate $l_{s}$ of Ge at 300 K. Our obtained
$l_{s0}$ of Ge is 2.4 $\mu$m, which is within the experimental range
from less than 0.5 $\mu$m to 34 $\mu$m,\citep{hamaya2018spin,rinaldi2016determination,zucchetti2019doping,yamada2021experimental}
but much shorter than 20 $\mu$m from Ref. \citenum{zucchetti2022electric}
and 34 $\mu$m from Ref. \citenum{zucchetti2019doping}. We find that
if theoretical $l_{s0}$ is tuned 20 $\mu$m to by multiplying a scaling
factor to the scattering term $L^{C}$, our theoretical values of
$l_{s}(E_{x})$ are in agreement with experimental data, as illustrated
in Fig. \ref{fig:ge}.

\subsection*{Appendix K: The derivation of Eq. \ref{eq:tausE_DP} from 1-RR simulation}

Within a minimum 1-RR simulation, for $\tau_{sx}(E_{x})$, we set
$V^{KR(L)}$ as
\begin{align}
V^{KR}= & \{U_{sx}^{tR},-L^{E}U_{sx}^{tR}\},\\
V^{KL}= & \{U_{sx}^{tL},-L^{E,\dagger}U_{sx}^{tL}\}.
\end{align}
Here we considered the fact that it is unnecessary to include $L^{E0}U_{sx}^{tR(L)}$
as $U_{sx}^{tR(L)}$ is eigenvector of $L^{E0}$. With such $V^{KR(L)}$,
we have
\begin{align}
I^{K}= & \left[\begin{array}{cc}
1 & -L_{1}^{E,xx}\\
-L_{1}^{E,xx} & L_{2}^{E,xx}
\end{array}\right],\\
-L^{E0K}= & \left[\begin{array}{cc}
\Gamma_{s0}, & U_{sx}^{tL,\dagger}L^{E0}L^{E}U_{sx}^{tR}\\
U_{sx}^{tL,\dagger}L^{E}L^{E0}U_{sx}^{tR}, & -U_{sx}^{tL,\dagger}L^{E}L^{E0}L^{E}U_{sx}^{tR}
\end{array}\right],\\
\approx & \left[\begin{array}{cc}
\Gamma_{s0}, & -\tau^{-1}L_{1}^{E,xx}\\
-\Gamma_{s0}L_{1}^{E,xx}, & -U_{sx}^{tL,\dagger}L^{E}L^{E0}L^{E}U_{sx}^{tR}
\end{array}\right]\\
-L^{EK}= & \left[\begin{array}{cc}
-L_{1}^{E,xx} & L_{2}^{E,xx}\\
L_{2}^{E,xx} & -L_{3}^{E,xx}
\end{array}\right],\\
L_{n}^{E,xx}= & U_{sx}^{tL,\dagger}\left(L^{E}\right)^{n}U_{sx}^{tR}.
\end{align}

Typically, it is found that $L_{1}^{E,xx}$ and $L_{3}^{E,xx}$ are
negligible compared with matrix elements, so that we approximate them
as 0. Define
\begin{align}
\tau^{-1}= & \frac{-U_{sx}^{tL,\dagger}L^{E}L^{E0}L^{E}U_{sx}^{tR}}{L_{2}^{E,xx}},\\
\left\langle \Omega_{E,x}^{2}\right\rangle = & -L_{2}^{E,xx},
\end{align}
where $\tau$ is an effective lifetime comparable to carrier lifetime.
we then have
\begin{align}
I^{K}\approx & \left[\begin{array}{cc}
1 & 0\\
0 & -\left\langle \Omega_{E,x}^{2}\right\rangle 
\end{array}\right],\\
-L^{E0K}\approx & \left[\begin{array}{cc}
\Gamma_{s0} & 0\\
0 & -\tau^{-1}\left\langle \Omega_{E,x}^{2}\right\rangle 
\end{array}\right]\nonumber \\
-L^{EK}\approx & \left[\begin{array}{cc}
0 & -\left\langle \Omega_{E,x}^{2}\right\rangle \\
-\left\langle \Omega_{E,x}^{2}\right\rangle  & 0
\end{array}\right].
\end{align}

Without external magnetic field, zero-electric-field spin relaxation
rate $\Gamma_{s,x}^{0}$ should be a real value and $\Gamma_{s,x}^{0}$=$\tau_{sx}^{-1}(E_{x}=0)$.
With the above matrices, by solving the reduced EVP Eq. \ref{eq:RR_EVP_taus},
we obtain electric-field-dependent spin relaxation rates
\begin{align}
\Gamma_{s\pm}(E_{x})= & \frac{\tau^{-1}+\tau_{sx}^{-1}(E_{x}=0)}{2}\nonumber \\
 & \pm\frac{\sqrt{(\tau^{-1}-\tau_{sx}^{-1}(E_{x}=0))^{2}-4\left\langle \Omega_{E,x}^{2}\right\rangle }}{2}.
\end{align}

In our studied $E$ range, $(4\left\langle \Omega_{E,x}^{2}\right\rangle )^{1/2}$$\ll\tau^{-1}-\tau_{sx}^{-1}(E_{x}=0)$
is satisfied. Therefore, the slower (spin) relaxation rate is approximately
\begin{align}
\Gamma_{s-}(E_{x})\approx & \tau_{sx}^{-1}(E_{x}=0)\nonumber \\
 & +(\tau^{-1}-\tau_{sx}^{-1}(E_{x}=0))^{-1}\left\langle \Omega_{E,x}^{2}\right\rangle .
\end{align}

In many systems including those studied here, $\tau_{sx}^{-1}(E_{x}=0)$$\ll\tau^{-1}$
is satisfied. Therefore,
\begin{align}
\Gamma_{s-}(E_{x})\approx & \tau_{sx}^{-1}(E_{x}=0)+\tau\left\langle \Omega_{E,x}^{2}\right\rangle ,
\end{align}
which is exactly Eq. \ref{eq:tausE_DP}.

Finally, we show that $\left\langle \Omega_{E,\alpha}^{2}\right\rangle $
is often negligible when spin precession is absent. In such cases,
spin relaxation is dominated by EY mechanism. Again, we suppose spins
are highly polarized along $\alpha$ direction or there are two Kramers
degenerate bands with one spin-up and the other spin-down. In these
cases, spin matrix (in Bloch basis) $s_{\alpha}$ is approximately
$k$-independent. Therefore, we have
\begin{align}
U_{sx}^{tR}\approx & \rho_{k}^{s_{\alpha}}=c^{\rho}\frac{df_{k}}{d\epsilon}s_{\alpha},\\
U_{sx}^{tL}\approx & \varrho^{s_{\alpha}}=c^{\varrho}s_{\alpha},
\end{align}
where $c^{\rho}$ and $c^{\varrho}$ are real constants and their
choices must satisfy $\varrho^{s_{\alpha}}\rho^{s_{\alpha}}$=1.

We also suppose $\xi$-part of $L^{E}$ is negligible, which is satisfied
in one-band systems and in many systems if smooth basis functions
are chosen. Thus, we can approximate $L^{E}\approx\frac{eE_{x}}{\hbar}\frac{d}{dk_{x}}$.
From integration by parts, we can easily prove that $\left\langle \Omega_{E,x}^{2}\right\rangle $=$-L_{2}^{E,xx}$$\approx$0.

\subsection*{Appendix L: Obtaining Eq. \ref{eq:BE} from approximate 0-RR simulation}

Below we will show that in a given two-band system, spin relaxation
in the case of a non-zero $E_{x}$ and ${\bf B}$=0 is similar to
spin relaxation in another case of $E_{x}$=0 and ${\bf B}$ equal
to ${\bf B}^{E_{x}}$ defined in Eq. \ref{eq:BE}.

Consider a two-band system, whose bands are originally Kramers degenerate
and splitted by spin-orbit fields ${\bf B}_{k}^{\mathrm{soc}}$ and
external magnetic field ${\bf B}$, the system Hamiltonian reads
\begin{align}
H_{k}= & \epsilon_{k}+H_{k}^{\mathrm{soc}}+H^{\mathrm{sZ}},\\
H_{k}^{\mathrm{soc}}= & \frac{\mu_{B}g_{0}\hbar}{2}{\bf B}_{k}^{\mathrm{soc}}\cdot\sigma,\\
H^{\mathrm{sZ}}= & \frac{\mu_{B}g_{0}\hbar}{2}{\bf B}\cdot\sigma.
\end{align}

In this case, we have (with original eigenbasis where bands are Kramers
degenerate)
\begin{align}
\rho_{k}^{s_{\alpha}}= & c^{\rho}\frac{df_{k}}{d\epsilon}\sigma_{\alpha},\\
\varrho_{k}^{s_{\alpha}}= & c^{\varrho}\sigma_{\alpha},
\end{align}
with $c^{\varrho}$ and $c^{\rho}$ real constants and $\left\langle \varrho^{s}|\rho^{s}\right\rangle $=1.
Therefore, we have (defining $L^{e0}$=$L^{e}({\bf B}=0)$)
\begin{align}
L^{e0}\rho^{s_{\alpha}}\approx & \frac{-i}{\hbar}\left[H_{k}^{\mathrm{soc}},\rho_{k}^{s_{\alpha}}\right],\\
L^{e0}\varrho^{s_{\alpha}}\approx & \frac{-i}{\hbar}\left[H_{k}^{\mathrm{soc}},\varrho_{k}^{s_{\alpha}}\right]
\end{align}

Within the 0-RR simulation, $V^{KR(L)}$ is set as
\begin{align}
V^{KR(L)}= & \{U_{sx}^{tR(L)},U_{sy}^{tR(L)},U_{sz}^{tR(L)}\},
\end{align}
with $U_{s\alpha}^{tR(L)}$ zero-field eigenvector corresponding to
the eigenvalue $\tau_{s0\alpha}^{-1}$. $\tau_{s0\alpha}$ is zero-field
spin lifetime.

If the scattering is relatively strong, which is usually true, we
have $\tau L^{e}\rho^{s_{\alpha}}$$\ll\rho^{s_{\alpha}}$ and $\tau L^{e}\varrho^{s_{\alpha}}$$\ll\varrho^{s_{\alpha}}$.
According to our prior paper (Ref. \citenum{xu2025predicting}), we
approximately have
\begin{align}
U_{s\alpha}^{tR}\approx & (1+\tau L^{e0})\rho^{s_{\alpha}},\\
U_{s\alpha}^{tL}\approx & (1-\tau L^{e0})\varrho^{s_{\alpha}}.
\end{align}

With the above $V^{KR(L)}$, the reduced EVP at $E_{x}$=0 is
\begin{align}
-L^{E0K}Y^{R}= & Y^{R}\Gamma_{s}^{E0},
\end{align}
where
\begin{align}
-L^{E0K}= & \left[\begin{array}{ccc}
\tau_{s0x}^{-1}-L^{B,xx}, & -L^{B,xy}, & -L^{B,xz}\\
-L^{B,yx}, & \tau_{s0y}^{-1}-L^{B,yy}, & -L^{B,yz}\\
-L^{B,zx}, & -L^{B,zy}, & \tau_{s0z}^{-1}-L^{B,zz}
\end{array}\right],\\
L^{B,\alpha\beta}= & U_{s\alpha}^{tL,\dagger}\frac{-i}{\hbar}\left[H^{\mathrm{sZ}},U_{s\beta}^{tR}\right].
\end{align}

Since $\tau L^{e}\rho^{s_{\alpha}}$$\ll\rho^{s_{\alpha}}$ and $\tau L^{e}\varrho^{s_{\alpha}}$$\ll\varrho^{s_{\alpha}}$,
we have
\begin{align}
L^{B,\alpha\beta}\approx & \mathrm{Tr}\int dk\varrho_{k}^{s_{\alpha}}\frac{-i}{\hbar}\left[H^{\mathrm{sZ}},\rho_{k}^{s_{\beta}}\right]\nonumber \\
= & \mathrm{Tr}\int dk\varrho_{k}^{s_{\alpha}}\frac{-i}{\hbar}\left[\frac{\mu_{B}g_{0}\hbar}{2}{\bf B}\cdot\sigma,\rho_{k}^{s_{\beta}}\right]\nonumber \\
= & \varepsilon_{\gamma\beta\alpha}\mu_{B}g_{0}B_{\gamma}^{\mathrm{ext}}.
\end{align}

The derivation uses the relation $[\sigma_{\gamma},\sigma_{\beta}]=2i\varepsilon_{\gamma\beta\alpha}\sigma_{\alpha}$.

Now, we consider the system at $E_{x}$$\neq$0 but ${\bf B}$=0.
The reduced EVP in this case is
\begin{align}
-L^{B0K}Y^{R}= & Y^{R}\Gamma_{s}^{B0},
\end{align}
where
\begin{align}
-L^{B0K}= & \left[\begin{array}{ccc}
\tau_{s0x}^{-1}-L_{1}^{E,xx}, & -L_{1}^{E,xy}, & -L_{1}^{E,xz}\\
-L_{1}^{E,yx}, & \tau_{s0y}^{-1}-L_{1}^{E,yy}, & -L_{1}^{E,yz}\\
-L_{1}^{E,zx}, & -L_{1}^{E,zy}, & \tau_{s0z}^{-1}-L_{1}^{E,zz}
\end{array}\right],\\
L_{1}^{E,\alpha\beta}= & U_{s\alpha}^{tL,\dagger}L^{E}U_{s\beta}^{tR}.
\end{align}

$L_{1}^{E,\alpha\beta}$ can be expressed as the sum of the following
four terms (note that $L^{e,\dagger}$=$-L^{e}$)
\begin{align}
L_{1}^{E,\alpha\beta}= & \left\langle \varrho^{s_{\alpha}}\right|L^{E}\left|\rho^{s_{\beta}}\right\rangle +\nonumber \\
 & \left\langle \varrho^{s_{\alpha}}\right|L^{E}\left|\tau L^{e}\rho^{s_{\beta}}\right\rangle +\nonumber \\
 & \left\langle -\tau L^{e}\varrho^{s_{\alpha}}\right|L^{E}\left|\rho^{s_{\beta}}\right\rangle +\nonumber \\
 & \left\langle -\tau L^{e}\varrho^{s_{\alpha}}\right|L^{E}\left|\tau L^{e}\rho^{s_{\beta}}\right\rangle .
\end{align}

Since $\tau L^{e}\rho^{s_{\alpha}}$$\ll\rho^{s_{\alpha}}$ and $\tau L^{e}\varrho^{s_{\alpha}}$$\ll\varrho^{s_{\alpha}}$,
the 4th term is negligible.

Suppose $\xi$-part of $L^{E}$ is negligible, which is satisfied
in many systems if smooth basis functions are chosen. Thus, we can
approximate $L^{E}\approx\frac{eE_{x}}{\hbar}\frac{d}{dk_{x}}$. From
integration by parts, the 1st and 2nd terms of $L_{1}^{E,\alpha\beta}$
- $\left\langle \varrho^{s_{\alpha}}\right|L^{E}\left|\rho^{s_{\beta}}\right\rangle $
and $\left\langle \varrho^{s_{\alpha}}\right|L^{E}\left|\tau L^{e}\rho^{s_{\beta}}\right\rangle $
are both zero. Moreover, from integration by parts, the 3rd term of
$L_{1}^{E,\alpha\beta}$ reads
\begin{align}
 & \left\langle -\tau L^{e}\varrho^{s_{\alpha}}\right|L^{E}\left|\rho^{s_{\beta}}\right\rangle \nonumber \\
\approx & -\frac{e}{\hbar}E_{x}\mathrm{Tr}\int dk\rho_{k}^{s_{\beta}}\frac{d}{dk_{x}}\left(-\tau L_{k}^{e}\varrho_{k}^{s_{\alpha}}\right)\nonumber \\
= & \tau\frac{-i}{\hbar}\frac{e}{\hbar}E_{x}\mathrm{Tr}\int dk\rho_{k}^{s_{\beta}}\left[\frac{dH_{k}^{\mathrm{soc}}}{dk_{k}},\varrho_{k}^{s_{\alpha}}\right]\nonumber \\
= & \tau\frac{-i}{\hbar}\frac{e}{\hbar}E_{x}\frac{\mu_{B}g_{0}\hbar}{2}\mathrm{Tr}\int dk\rho_{k}^{s_{\beta}}\left[\frac{d{\bf B}_{k}^{\mathrm{soc}}}{dk_{x}}\cdot\sigma,\varrho_{k}^{s_{\alpha}}\right]\nonumber \\
= & -\varepsilon_{\gamma\beta\alpha}\frac{e\mu_{B}g_{0}}{\hbar}E_{x}\tau\left\langle \varrho^{s_{\beta}}\right|\frac{dB_{k\gamma}^{\mathrm{soc}}}{dk_{x}}\left|\rho^{s_{\beta}}\right\rangle .
\end{align}

The derivation considers $\varepsilon_{\gamma\alpha\beta}=-\varepsilon_{\gamma\beta\alpha}$.
$\left\langle \varrho^{s_{\beta}}\right|dB_{k\gamma}^{\mathrm{soc}}/dk_{x}\left|\rho^{s_{\beta}}\right\rangle $
can be understood as Fermi-surface average of $dB_{k\gamma}^{\mathrm{soc}}/dk_{x}$
- $\left\langle dB_{k\gamma}^{\mathrm{soc}}/dk_{x}\right\rangle $.
Therefore,
\begin{align}
L_{1}^{E,\alpha\beta}\approx & \left\langle -\tau L^{e}\varrho^{s_{\alpha}}\right|L^{E}\left|\rho^{s_{\beta}}\right\rangle \nonumber \\
\approx & -\varepsilon_{\gamma\beta\alpha}\frac{e\mu_{B}g_{0}}{\hbar}E_{x}\tau\left\langle \frac{dB_{k\gamma}^{\mathrm{soc}}}{dk_{x}}\right\rangle ,
\end{align}
which is the same as $L^{B,\alpha\beta}$ if $B_{\gamma}^{\mathrm{ext}}$=$-e\hbar^{-1}E_{x}\tau\left\langle dB_{k\gamma}^{\mathrm{soc}}/dk_{x}\right\rangle $.
This indicates that an electric field $E_{x}$ indeed generates an
effective magnetic field as defined in Eq. \ref{eq:BE}.

\bibliography{ref}

\end{document}